\documentclass[paper]{JHEP3}
\pdfoutput=1
\usepackage{amsmath,amssymb,amsthm,amscd,graphicx}
\input epsf.sty

\theoremstyle{definition}

\newcommand{\sn}{{\rm{sn}}}
\newcommand{\tn}{{\rm{tn}}}

\newcommand{\CI}{{\cal I}}

\newcommand{\CN}{{\cal N}}
\newcommand{\CO}{{\cal O}}

\newcommand{\CS}{{\cal S}}

\def\IR{{\mathbb R}}

\def\IP{{\mathbb P}}

\def\IN{{\mathbb N}}

\newcommand{\re}{{\rm e}}
\newcommand{\ri}{{\rm i}}
\newcommand{\rd}{{\rm d}}

\newcommand{\be}{\begin{equation}}
\newcommand{\ee}{\end{equation}}
\newcommand{\ba}{\begin{aligned}}
\newcommand{\ea}{\end{aligned}}
\newcommand{\ben}{\begin{eqnarray}\displaystyle}
\newcommand{\een}{\end{eqnarray}}

\newcommand{\sectiono}[1]{\section{#1}\setcounter{equation}{0}}

\newdimen\tableauside\tableauside=1.0ex
\newdimen\tableaurule\tableaurule=0.4pt
\newdimen\tableaustep
\def\phantomhrule#1{\hbox{\vbox to0pt{\hrule height\tableaurule width#1\vss}}}
\def\phantomvrule#1{\vbox{\hbox to0pt{\vrule width\tableaurule height#1\hss}}}
\def\sqr{\vbox{%
  \phantomhrule\tableaustep
  \hbox{\phantomvrule\tableaustep\kern\tableaustep\phantomvrule\tableaustep}%
  \hbox{\vbox{\phantomhrule\tableauside}\kern-\tableaurule}}}
\def\squares#1{\hbox{\count0=#1\noindent\loop\sqr
  \advance\count0 by-1 \ifnum\count0>0\repeat}}
\def\tableau#1{\vcenter{\offinterlineskip
  \tableaustep=\tableauside\advance\tableaustep by-\tableaurule
  \kern\normallineskip\hbox
    {\kern\normallineskip\vbox
      {\gettableau#1 0 }%
     \kern\normallineskip\kern\tableaurule}%
  \kern\normallineskip\kern\tableaurule}}
\def\gettableau#1{\ifnum#1=0\let\next=\null\else
\squares{#1}\let\next=\gettableau\fi\next}

\newcommand{\figref}[1]{Fig.~\protect\ref{#1}}

\title{Instanton effects and quantum spectral curves}

\author{
Johan K\"all\'en and Marcos Mari\~no \\
D\'epartement de Physique Th\'eorique et Section de Math\'ematiques,\\
Universit\'e de Gen\`eve, Gen\`eve, CH-1211 Switzerland\\
\\
\email{johan.kallen@unige.ch, marcos.marino@unige.ch }
}

\abstract{We study a spectral problem associated to the quantization of a spectral curve arising in local mirror symmetry. 
The perturbative WKB quantization condition is determined by 
the quantum periods, or equivalently by the refined topological string in the Nekrasov--Shatashvili (NS) limit. 
We show that the information encoded in the quantum periods is radically insufficient 
to determine the spectrum: there is an infinite series of instanton corrections, which are non-perturbative in $\hbar$, and lead to an exact WKB quantization condition. Moreover, we conjecture the precise form of the instanton corrections: they are determined by the standard or un-refined topological string free energy, and we test our conjecture successfully against numerical calculations of the spectrum. This suggests that the non-perturbative sector of the NS refined topological string contains 
information about the standard topological string. As an application of the WKB quantization condition, we explain some recent observations relating membrane instanton corrections in ABJM theory to the refined topological string.
}

\begin{document}

\sectiono{Introduction}

This paper is motivated by two different, but related problems. The first problem is the non-perturbative structure of 
topological string theory. Topological strings, like many other string models, are only defined perturbatively, and it is natural to 
ask whether one can define them non-perturbatively or find new non-perturbative sectors. In the last years, there have been many 
different proposals addressing this problem, but none of them seems to be conclusive. A {\it bona fide} 
non-perturbative definition must be based on a manifestly well-defined quantity, at least for a certain range of the relevant parameters of the model. 
This quantity should have an asymptotic expansion, for small values of the string coupling constant, 
which reproduces the original perturbative expansion. The asymptotic expansion and the perturbative expansion 
can only differ in quantities which are non-analytic at the origin (like for example instanton effects). Of course, one might find 
different non-perturbative definitions of the same quantity, all of them differing in non-analytic terms. This is for example what happens 
in two-dimensional gravity \cite{dfgzj}. In some cases, a reasonable physical criterium might single out one non-perturbative definition. 

In this paper we will analyze the non-perturbative structure of refined topological strings in the Nekrasov--Shatashvili (NS) limit \cite{ns}. 
This theory depends on the Calabi--Yau (CY) moduli and on a string coupling constant which is usually denoted by $\hbar$. 
In the original proposal of \cite{ns}, this refined string was related, for some special geometries, to quantum integrable systems. 
It was later pointed out in \cite{mm,acdkv} that the perturbative free energy of the NS topological string can be computed by 
quantizing a spectral curve given by the mirror Calabi--Yau or a limit thereof. Using the quantum spectral curve one can 
construct quantum periods, depending on $\hbar$, which define the free energy by an $\hbar$-deformed version of special geometry. 
The quantum periods have a nice interpretation in terms of one-dimensional Quantum Mechanics: the quantized spectral curve 
defines a spectral problem, and the quantum periods are quantum-corrected WKB periods, which 
lead to a perturbative quantization condition, at all orders in $\hbar$. 

In this paper we study a spectral problem appearing in the quantization of the curve describing 
the mirror of local $\IP^1 \times \IP^1$. We show 
that the perturbative quantum periods are radically insufficient to solve the spectral problem: there is an infinite series of 
non-perturbative corrections in $\hbar$, of the instanton type\footnote{The instanton 
effects we study in this paper are not related to the  gauge-theory instantons appearing in Nekrasov's calculation of the topological 
string free energy \cite{n-inst}. These gauge-theory instantons are non-perturbative in $\alpha'$ and are 
already incorporated in the perturbative WKB periods.}. 
This is a well-known phenomenon in ordinary Quantum Mechanics. For example, in the double-well potential, 
the standard WKB quantization condition is insufficient to determine the spectrum, even after including 
all perturbative corrections in $\hbar$: one should also take into account instantons tunneling between the two vacua, 
and including these leads to a non-perturbative, exact quantization condition \cite{zj,zjj}. 
In the problem at hand a similar phenomena 
occurs, but it is even more dramatic: for some values of $\hbar$, the quantization condition based on the quantum periods leads to an unphysical 
divergent expression. Instanton corrections are needed to cure the divergence. 

A first-principle calculation of these instanton corrections is difficult, but we conjecture their precise form: they involve the 
{\it standard} (i.e. un-refined) topological string free energy. In particular, we write down an 
{\it exact} WKB quantization condition involving both the perturbative quantum periods and the instanton corrections. 
We perform a very precise test of this conjecture by comparing the exact quantization condition to the numerical calculation of the spectrum. 
The agreement is excellent.

Our conjecture gives a novel realization of the Gopakumar--Vafa invariants of local $\IP^1 \times \IP^1$ in terms of a spectral problem. 
More generally, it suggests that the spectral problems associated to quantum spectral curves in local mirror symmetry involve 
the standard topological string, non-perturbatively. It also suggests that we should {\it define} non-perturbative topological strings on local CYs through a well-defined spectral problem 
associated to the quantization of the mirror curve. For example, for some local CY geometries, the results of \cite{ns} provide a description of the refined topological string 
in terms of a quantum integrable system, and this should lead naturally to the sought-for spectral problem. If the structure we find in our particular example 
generalizes to other cases, this definition leads in a single strike to the refined NS string (as the perturbative sector) and the conventional topological string 
(as the non-perturbative sector). Notice that this non-perturbative definition, in the example of local $\IP^1 \times \IP^1$, satisfies the above two criteria: it is based on a well-defined quantity (the spectrum), and it reproduces perturbatively the quantum B-period (which is the derivative of the free energy). Of course, 
the most interesting aspect of this definition is the appearance of the un-refined topological string free energy in the non-perturbative sector. 

The second motivation for our work comes from ABJM theory \cite{abjm}. In \cite{kwy}, 
the partition function of this theory on a three-sphere was obtained by using localization techniques, and written as a matrix integral. 
Its full 't Hooft $1/N$ expansion was obtained in \cite{dmp} by using large $N$ techniques. In the paper \cite{mp}, this matrix integral was written as the thermal partition function of 
an ideal, one-dimensional Fermi gas. One advantage of this approach, as compared to the standard large $N$ techniques, 
is that one can also compute non-perturbative effects due to membrane instantons, which go beyond the 't Hooft expansion. In \cite{hmmo}, building on previous work \cite{hmo,hmo1,cm,hmo2}, it was conjectured that these non-perturbative effects are encoded in the quantum periods of local $\IP^1\times \IP^1$. 
In this paper we prove this conjecture to a large extent. The basic idea is simple: the spectral problem associated to the one-particle Hamiltonian of the Fermi gas is nothing but 
the spectral problem studied in this paper, i.e. it is a specialization of the spectral problem 
appearing in the quantization of the spectral curve of local $\IP^1 \times \IP^1$. The WKB analysis of the ABJM 
spectral problem leads immediately to the connection with these quantum periods. The grand potential of 
ABJM theory can be computed once we know the spectrum from the WKB quantization condition, 
and this makes it possible to derive many aspects of this grand potential which were conjectured in \cite{hmmo}. 

Many of our results on the WKB approach to the spectral problem are dual to the results on the grand potential  of ABJM theory. For example, the fact that quantum periods are divergent for some values of $\hbar$ and should get non-perturbative corrections which cancel these divergences is a dual version of the HMO cancellation mechanism of \cite{hmo1}. Our conjecture on the instanton corrections to the spectral problem was motivated to a large extent by the known worldsheet instanton corrections to the grand potential of ABJM theory. 

The organization of this paper is as follows: in section 2 we present the spectral problem we will focus on. In section 3 we do a WKB analysis of this problem. We first show that quantum periods are insufficient, we conjecture the form of the instanton corrections, and we perform a detailed test against the numerical calculation of the spectrum. In section 4 we derive from the results in section 3 the structure of the grand potential of ABJM theory, proving in this way some of the conjectures in \cite{hmmo}. Finally, in section 5 we state our conclusions and directions for further research. An Appendix contains some results on Mellin transforms which are used in section 4.

\sectiono{The spectral problem}

We will consider a spectral problem arising in the quantization of the spectral curve describing the local CY known as local $\IP^1 \times \IP^1$. This is a spectral 
problem for a difference equation, with an appropriate and natural choice of analyticity and boundary conditions. The resulting problem 
can be equivalently formulated in terms of an integral equation which plays a crucial role in the Fermi gas approach to ABJM theory \cite{mp}. In this section we 
will first consider the integral equation formulation, and then the formulation in terms of a difference equation. 

Let us consider the following integral kernel
\be
\label{rhok}
\rho(x_1, x_2)={1\over 2 \pi k} {1\over \left( 2 \cosh  {x_1 \over 2}  \right)^{1/2} }  {1\over \left( 2 \cosh {x_2  \over 2} \right)^{1/2} } {1\over 
2 \cosh\left( {x_1 - x_2\over 2 k} \right)}. 
\ee
Here, $k$ is a real parameter. This is a particular case of the family of kernels studied in \cite{zamo,tw}. The spectral problem associated to this integral kernel is 
\be
\label{int-eq}
\int_{-\infty}^\infty \rho (x_1, x_2) \phi (x_2) \rd x_2= \re^{-E} \phi(x_1).
\ee
The kernel (\ref{rhok}) defines a non-negative, Hermitian, Hilbert--Schmidt operator, therefore it has a discrete, positive spectrum 
\be
0<E_0 < E_1 < E_2 < \cdots. 
\ee
It is easy to reformulate (\ref{int-eq}) as a spectral problem for a difference equation. One way to do this is to consider the operator $\hat \rho$ defined by \cite{mp}
 \be
\langle x| \hat  \rho | x' \rangle=\rho(x, x'). 
 \ee
 This operator can be written as
\be
\label{rhopq}
\hat \rho=\re^{-{1\over 2} U(\hat x)} \re^{-T(\hat p)} \re^{-{1\over 2} U(\hat x)}.
\ee
In this equation, $\hat x, \hat p$ are canonically conjugate operators, 
\be
[\hat x, \hat p]=\ri \hbar, 
\ee
where
\be
\label{hbar}
\hbar=2 \pi  k, 
\ee
and
\be
\label{ut}
U(x)=\log \left( 2 \cosh {x\over 2} \right), \qquad T(p)=\log\left( 2\cosh  {p \over 2} \right). 
\ee
In this paper we will use $\hbar$ and $k$ interchangeably. The spectral problem (\ref{int-eq}) can now be written as
\be
\hat \rho |\phi\rangle= \re^{-E} |\phi\rangle. 
\ee
Let us now define 
\be
|\psi \rangle =\re^{{1\over 2} U(\hat x)} |\phi \rangle.
\ee
It follows that
\be
\re^{ U(\hat x)} \re^{T(\hat p)}|\psi \rangle = \re^{E} |\psi\rangle
\ee
or, equivalently, in the coordinate representation,
\be
\label{diff-eq}
\psi \left( x +\ri \pi k\right) +  \psi \left( x-\ri \pi k\right) = {\re^E \over 2 \cosh \left( {x\over 2} \right)} \psi(x). 
\ee
This difference equation is only equivalent to the original problem (\ref{int-eq}) provided some analyticity and boundary conditions are imposed on the function $\psi(x)$. 
Following \cite{tw}, let us denote by $\CS_a$ the strip in the complex $x$-plane defined by 
\be
\left|{\rm Im}(x)\right|<a. 
\ee
Let us also denote by $A\left(\CS_{a} \right)$ those functions $g$ which are bounded and analytic in the strip, continuous on its closure, and for which 
$g(x+ \ri y) \to 0$ as $x\rightarrow \pm \infty$ through real values, when $y \in \IR$ is fixed and satisfies $|y|<a$. 
It can be seen, by using the results in \cite{tw}, that the equivalence of (\ref{diff-eq}) and (\ref{int-eq}) requires that $\psi(x)$ belongs to the space $A\left(\CS_{\pi k} \right)$. 

The operator $\hat \rho$ can be used to define a quantum Hamiltonian in the usual way \cite{mp}, 
\be
\hat \rho =\re^{-\hat H}, 
\ee
whose classical limit is simply 
\be
\label{class-H}
H_{\rm cl}(x, p)= T(p) + U(x). 
\ee
In \cite{mp} it was noticed that the curve 
\be
\label{class-curve}
\exp \left( T(p) + U(x) \right)=\re^E,
\ee
defining the classical limit of the spectral problem, is a specialization of the curve describing the mirror of the 
Calabi--Yau known as local $\IP^1 \times \IP^1$. Let us write this curve as in \cite{acdkv}
\be
\label{localp1p1}
\re^u + z_1 \re^{-u}+ \re^v + z_2 \re^{-v}= 1, 
\ee
where $u,v$ are complex coordinates. Then, the curve (\ref{class-curve}) can be seen to be equal to the curve (\ref{localp1p1}) after the change of variables
\be
\label{cv}
u={x+p \over 2} -E, \qquad v={x-p \over 2}-E
\ee
and the specialization 
\be
\label{zz}
z_1= z_2=z,
\ee
where we have denoted, for convenience,
\be
z= \re^{-2E}. 
\ee
Notice that the above change of variables is essentially a canonical transformation, since it preserves the symplectic form, up to an overall constant, 
\be
\rd u \wedge \rd v =-{1\over 2} \rd x \wedge \rd p. 
\ee
The curve (\ref{localp1p1}) can be quantized, leading to a {\it quantum spectral curve}. One simply promotes $u,v$ to quantum operators $\hat u$, $\hat v$ satisfying canonical commutation 
relations. The equation satisfied by wavefunctions is  
\be
\label{qsc}
\left( \re^{\hat u} + z_1 \re^{-\hat u}+ \re^{\hat v} + z_2 \re^{-\hat v}- 1\right) |\psi \rangle=0. 
\ee
We can now regard (\ref{diff-eq}) as a particular case of (\ref{qsc}) by promoting the classical change of variables (\ref{cv}) to a quantum one, 
\be
\hat u= {\hat x+\hat p \over 2}+{\ri \pi k \over 4}  -E, \qquad \hat v={\hat x-\hat p \over 2}-{\ri \pi k \over 4} -E, 
\ee
while the specialization (\ref{zz}) has now the quantum correction\footnote{This change of variables was previously observed by Kazumi Okuyama.}
\be
\label{zz-q}
z_1=q^{1/2} z, \qquad z_2=q^{-1/2}z, 
\ee
where
\be
\label{qh}
q=\re^{\ri \hbar \over 2}= \re^{\pi \ri k}.
\ee
In terms of $\hbar$ as defined in (\ref{hbar}), we have
\be
[\hat v, \hat u]= {\ri \hbar \over 2}. 
\ee
We conclude that the results for the quantum spectral curve (\ref{qsc}) obtained for example in \cite{acdkv,hmmo} 
can be specialized to study the spectral problem (\ref{int-eq}) and (\ref{diff-eq}). 

The difference equation (\ref{diff-eq}) has the structure of Baxter's TQ equation, which determines 
the spectrum of a quantum integrable system and can be regarded as a quantization of the spectral curve of the classical system. This similarity is not surprising: 
as it is well-known, the curve (\ref{localp1p1}), describing the mirror of local $\IP^1 \times \IP^1$, can be regarded as a relativistic deformation of the 
spectral curve of the periodic Toda chain \cite{nekrasov}. The Baxter equations for Toda and relativistic Toda have been studied in \cite{pg} and \cite{sergeev}, respectively. 
Similar difference equations also appear in the study of $\CN=2$ supersymmetric gauge theories in the NS limit \cite{pogho,fmpp}. 
(\ref{diff-eq}) is also a close cousin of the difference equation studied in \cite{flz}, which has its origin in the integral equation of the 't Hooft model. 

Unfortunately, the spectral problem (\ref{int-eq}) does not seem to be exactly solvable, and one has to use numerical or approximate methods.

\sectiono{The exact WKB quantization condition}

\subsection{Perturbative WKB quantization}

We will now analyze the spectral problem (\ref{diff-eq}) by using the WKB method. At leading order in $\hbar$, the WKB method for bound states is the Bohr--Sommerfeld quantization condition. In this approximation, one calculates the classical volume of phase space as a function of the energy, ${\rm vol}_{0}(E)$ (here, the subscript $0$ means that we are working at zero order in the $\hbar$ expansion). The quantization condition says that this volume should be a half-integer multiple\footnote{The fact that this is a half-integer, and not an integer, can be shown by using a next-to-leading WKB analysis, as in \cite{pg}: the presence of an inverse square-root factor in the WKB wavefunction leads to an extra phase in going around a cut.} of the volume of an elementary cell in phase space, $2 \pi \hbar$, and one obtains
\be
\label{bs}
{\rm vol}_0 (E) = 2 \pi \hbar \left( n+{1\over 2}\right), \qquad n=0, 1,2, \cdots. 
\ee
The classical volume of phase space was already determined in \cite{mp}. It is given by a period integral on the curve (\ref{class-curve}). Since this is an elliptic curve, it has two periods, the $A$ and the $B$ periods. The spectral problem we are looking at involves the $B$ period, and we find
\be
{\rm vol}_0(E)= \oint_B \lambda,\qquad \lambda= p(x) \rd x, 
\ee
where $p(x)$ is obtained by solving (\ref{class-curve}). The calculation in \cite{mp} expresses this period in terms of a Meijer G-function
\be
 \label{volexact}
{\rm vol}_0(E)=\frac{\re^E}{\pi}  G_{3,3}^{2,3}\left(\frac{\re^{2E}}{16}\left|
\begin{array}{c}
 \frac{1}{2},\frac{1}{2},\frac{1}{2} \\
 0,0,-\frac{1}{2}
\end{array}
\right.\right)-4\pi^2 = 8 E^2 -{4  \pi^2 \over 3} + \CO\left( E\, \re^{-2E} \right).
\end{equation}
In the WKB method we are interested in large energies as compared to $\hbar$, i.e. in large quantum numbers. It is then useful to have a basis of classical periods 
of the curve which is appropriate for the $E \gg 1$ regime. Since the curve (\ref{class-curve}) is a specialization of the mirror of 
local $\IP^1 \times \IP^1$, the relevant periods 
are nothing but the large radius periods of this CY. Let us now review some basic facts about these periods. 

In the full CY (\ref{localp1p1}), i.e. for generic values of $z_1, z_2$, it is useful to consider two different $A$ periods and two different $B$ periods. 
The $A$ periods are given by
\be
\Pi_{A_I}(z) =\log z_I + \widetilde \Pi_A (z_1, z_2), \qquad I=1,2,
\ee
where
\be
\label{Aper}
\widetilde \Pi_A (z_1, z_2)= 2\sum_{k,l\ge 0, \atop (k,l)\not=(0,0)} { \Gamma(2k + 2l) \over \Gamma(1+k)^2 \Gamma(1+l)^2} z_1^k z_2^l =2z_1 + 2z_2 + 3 z_1^2 + 12 z_1 z_2 + 3 z_2^2 + \cdots
\ee
There are two independent $B$-periods, $\Pi_{B_I}(z_1, z_2)$, $I=1,2$, which are related by the exchange of $z_1$ and $z_2$, 
\be
\Pi_{B_2}(z_1, z_2)=\Pi_{B_1}(z_2, z_1). 
\ee
The $B_1$ period is given by 
\be
\label{B1per}
\Pi_{B_1}(z_1, z_2)=-{1\over 8}\left( \log^2 z_1  -2 \log z_1\log z_2 -\log^2 z_2 \right) +{1\over 2} \log z_2\,  \widetilde \Pi_A (z_1, z_2)+ {1\over 4}  \widetilde \Pi_B (z_1, z_2), 
\ee
where
\be
\label{Bper}
\ba
\widetilde \Pi_B (z_1, z_2) &= 8 \sum_{k,l\ge 0, \atop (k,l)\not=(0,0)} { \Gamma(2k + 2l) \over \Gamma(1+k)^2 \Gamma(1+l)^2} \left( \psi(2k+2l) -\psi(1+l)\right) z_1^k z_2^l \\
 &= 8 z_1+ 22 z_1^2 + 40 z_1 z_2 + 4z_2^2 +\cdots
 \ea
 \ee

In our spectral problem we have $z_1=z_2$ classically (see (\ref{zz})). In this limit, one has
\be
\widetilde \Pi_{A}(z)\equiv \widetilde \Pi_A(z,z)=\sum_{\ell \ge 1}  \widehat a_\ell^{(0)} z^\ell, \qquad 
\widetilde \Pi_{B}(z)\equiv \widetilde \Pi_B (z,z)=\sum_{\ell \ge 1} \widehat b_\ell^{(0)}  z^\ell, 
\ee
where
\be
\label{aobo}
\ba
\widehat a_\ell^{(0)} &={1\over  \ell} \left( {\Gamma \left( \ell+{1\over 2} \right) \over 
\Gamma({1\over 2}) \ell!} \right)^2 16^{\ell}~, \\
\widehat b_\ell^{(0)} &= {4 \over  \ell} \left( {\Gamma \left( \ell+{1\over 2} \right) \over 
\Gamma({1\over 2}) \ell!} \right)^2 16^{\ell} \left[ \psi\left(\ell+{1\over 2} \right) -\psi(\ell+1)+ 2 \log 2-{1\over 2\ell} \right] ~. 
\ea
\ee
The classical B-period becomes
\be
\Pi_B(z)\equiv \Pi_{B_{1,2}}(z,z)={1\over 4} \left( \log z\right)^2 + {1\over 2} \log z  \, \widetilde \Pi_A (z)+{1\over 4} \widetilde \Pi_B(z),
\ee
and one finds that the volume of phase space can be written in terms of this period as
\be
{\rm vol}_0(E)= 8 \Pi_B \left(\re^{-2E} \right)-{4 \pi^2 \over 3}= 8 E^2 -{4 \pi^2 \over 3} - 8 E \sum_{\ell\ge 1} \widehat a_\ell^{(0)}  \re^{-2 \ell E} 
+ 2 \sum_{\ell\ge1} \widehat b_\ell^{(0)} \re^{-2 \ell E}. 
\ee

It is well-known that the Bohr--Sommerfeld quantization condition has perturbative corrections in $\hbar$. These can be obtained in a straightforward way 
by solving the equations (\ref{diff-eq}) with a WKB ansatz, 
\be
\label{WKBpsi}
\psi(x, \hbar) = \exp \left( {1\over \hbar} S(x, \hbar)\right), 
\ee
where
\be
\label{WKBS}
S(x,\hbar)= \sum_{n\ge 0} S_n(x) \hbar^{n}, 
\ee
and interpreting $\partial_x S(x, \hbar) \rd x$ as a ``quantum" differential. The leading order approximation gives
\be
S_0'(x)= p(x) 
\ee
and reproduces the Bohr--Sommerfeld quantization condition. Using this quantum differential, we can define the perturbative, ``quantum" volume of phase space as  
\be
\label{p-vol}
{\rm vol}_{\rm p} (E; \hbar)= \oint_B \partial_x S(x, \hbar) \rd x. 
\ee
We could calculate these corrections directly in the equation (\ref{diff-eq}). However, it is more illuminating to obtain them as particular cases of the quantum corrections for the 
spectral curve (\ref{qsc}). Indeed, as explained in \cite{mm,acdkv} and reviewed in \cite{hmmo}, these corrections promote the classical periods $\Pi_{A_I}(z_1, z_2)$, $\Pi_{B_I}(z_1, z_2)$ to quantum A-periods 
\be
\Pi_{A_I}(z_1,z_2;\hbar)=\log z_I +\widetilde{\Pi}_{A}(z_1,z_2;\hbar), \quad I=1,2, 
\ee
and quantum B-periods $\Pi_{B_I} (z_1, z_2; \hbar)$, $I=1, 2$. As in the classical case, there are two of them, 
but they are related by the exchange of the moduli, 
\be
\Pi_{B_2}(z_1,z_2;\hbar)=\Pi_{B_1}(z_2,z_1;\hbar).
\ee
The quantum counterpart of (\ref{B1per}) is 
\be
\ba
\label{q-B}
\Pi_{B_1}(z_1, z_2;\hbar)&=-{1\over 8}\left( \log^2 z_1  -2 \log z_1\log z_2 -\log^2 z_2 \right) +{1\over 2} \log z_2\, \widetilde \Pi_A (z_1, z_2;\hbar)\\
&+ {1\over 4} \widetilde \Pi_B (z_1, z_2;\hbar).
\ea
\ee
These quantum periods can be computed systematically in a power series in $z_{1,2}$ \cite{acdkv,hmmo}. One finds, to the very first orders, 
\be
\label{q-ABper}
\ba
\widetilde{\Pi}_{A}(z_1,z_2;\hbar)&=2(z_1+z_2)+3(z_1^2+z_2^2)+2(4+q+q^{-1})z_1z_2+\frac{20}{3}(z_1^3+z_2^3) \\
&\quad +2(16+6q+6q^{-1}+q^2+q^{-2})z_1 z_2(z_1+z_2)+{\cal O}(z_i^4), \\
\widetilde{\Pi}_{B}(z_1, z_2;\hbar)&=8 \left[\frac{q+1}{2(q-1)} \log q \right] z_1+4 \left[ 1+\frac{5 q^2 + 8 q + 5}{2(q^2-1)}\log q \right] z_1^2 
\\ 
& +8 \left[ 1+\frac{(1+q)^3}{2 q(q-1)}\log q \right]z_1z_2
+4 z_2^2+{\cal O}(z_i^3),
\ea
\ee
where $q$ is given in (\ref{qh}). 

Let us now come back to the problem of calculating (\ref{p-vol}). This is a quantum period for the spectral curve defined by (\ref{diff-eq}), but this curve is just a specialization of (\ref{qsc}) with the dictionary (\ref{zz-q}) and after a canonical transformation. Therefore, (\ref{p-vol}) should be a combination of the quantum periods of local $\IP^1 \times \IP^1$, specialized to the ``slice" (\ref{zz-q}). Let us denote
\be\label{q-ABperrelab}
\ba
\widetilde \Pi_{A}(z; 
\hbar)& \equiv \widetilde \Pi_A(q^{1/2}z,q^{-1/2}z; \hbar)=\sum_{\ell \ge 1}  \widehat a_\ell (\hbar) z^\ell, \\ 
\widetilde \Pi_{B}(z;\hbar)& \equiv {1\over 2} \left( \widetilde \Pi_B (q^{1/2}z,q^{-1/2}z; \hbar)+ \widetilde \Pi_B (q^{-1/2}z,q^{1/2}z; \hbar)\right) =\sum_{\ell \ge 1} \widehat b_\ell (\hbar)  z^\ell,
\ea
\ee
where $\widehat a_\ell (\hbar)$, $\widehat b_\ell (\hbar)$ have the $\hbar$-expansion, 
\be
\widehat a_\ell (\hbar)= \sum_{n=0}^\infty \widehat a_\ell ^{(n)} \hbar^{2n}, \qquad \widehat b_\ell (\hbar)= \sum_{n=0}^\infty \widehat b_\ell ^{(n)} \hbar^{2n}.
\ee
Requiring the combination of quantum periods to have the correct classical limit, and that only even powers of $\hbar$ appear, we find, 
\be
\ba
\label{pvol}
{\rm vol}_{\rm p} (E; \hbar)&= 4 \Pi_{B_1} ( q^{1/2} z, q^{-1/2} z; \hbar) + 4 \Pi_{B_2} ( q^{1/2} z, q^{-1/2} z; \hbar)  -{4 \pi^2 \over 3} -{\hbar^2 \over 12}\\
&= 8 E^2 -{4 \pi^2 \over 3} +{\hbar^2 \over 24} - 8 E \sum_{\ell\ge 1}  \widehat a_\ell (\hbar) \re^{-2 \ell E} + 2 \sum_{\ell\ge1} \widehat b_{\ell } (\hbar) \re^{-2 \ell E}. 
\ea
\ee
The third term in the last line of (\ref{pvol}) 
is an $E$-independent correction to the quantum period which was already computed in \cite{mp}. Notice that the perturbative quantum volume has an $\hbar$-expansion of the 
form, 
\be
\label{vol-p-exp}
{\rm vol}_{\rm p} (E)=\sum_{n\geq0} {\rm vol}_n(E)\hbar^{2n}~. 
\ee
The first term in this expansion is the function of $E$ given in (\ref{volexact}). By the WKB expansion of the wavefunction (\ref{WKBpsi}), (\ref{WKBS}), it is possible 
to find an exact expression for the first quantum correction, which reads
\be
\label{vol01}
{\rm vol}_1(E)=\frac{\re^{-E} \left((32 \, \re^{-2E}-1) E(k)-K(k)\right)}{6 \left(16 \,  \re^{-2E}-1 \right)},
\ee
where $K(k)$, $E(k)$ are elliptic integrals of the first and second kind, respectively, and their modulus is given by 
\be
k^2=1-\frac{\re^{2E}}{16 }~.
\ee
The derivation of this result is sketched in appendix \ref{qvolder}.

The equation (\ref{p-vol}) gives the full series of perturbative $\hbar$ corrections to the classical phase-space volume. The perturbatively 
exact quantization condition involves the quantum B-periods of the spectral curve, and it reads
\be
\label{ns-p}
{\rm vol}_{\rm p} (E;\hbar) = 2 \pi \hbar \left( n+{1\over 2}\right), \qquad n=0, 1,2, \cdots
\ee
We can now use (\ref{ns-p}) to compute the quantum-corrected spectrum. For example, we can use the explicit expressions (\ref{volexact}) and (\ref{vol01}) to compute the energies $E_n$ 
perturbatively, as a power series expansion around $\hbar=0$, i.e. 
\be
\label{enseries}
E^{\rm p}_n =\sum_{\ell=0}^\infty  E_{n,\ell} \hbar^\ell. 
\ee
The leading order term corresponds to the zero of ${\rm vol}_0(E)$, and as found in \cite{mp} this is, 
\be
E_{n,0}= 2\log 2. 
\ee
Plugging now the series (\ref{enseries}) in (\ref{ns-p}), we find 
\be
\label{ens}
 E_{n,1}=\frac{2n+1}{8}~,\quad E_{n,2}=-\frac{2n^2+2n+1}{128}. 
 \ee
 These values agree with a calculation starting directly from the density operator (\ref{rhopq}) \cite{hmo-un}.

\subsection{Non-perturbative WKB quantization}

As we have seen, 
the perturbative WKB quantization condition (\ref{ns-p}) makes it possible to compute the energies as a power series in $\hbar$. However, when we consider finite values of $\hbar$, we find 
a key problem: the coefficients $\widehat b_\ell (\hbar)$ appearing in the expansion (\ref{pvol}) 
{\it diverge} for any integer $k$, and lead to a non-sensical WKB expansion 
when $\hbar$ approaches $2 \pi$ times an integer. At the same time, there is no physical source for this divergence in the 
spectral problem itself: the eigenvalues for $E$ appearing in (\ref{diff-eq}) and in the associated integral equation (\ref{int-eq}) 
are perfectly well-defined for any real value of $\hbar$, 
and in particular for integer $k$. As a matter of fact they can be computed numerically, as we will see in the next subsection. We stress that the divergence problem 
in (\ref{ns-p}) is not an artifact of the large $E$ expansion used to obtain (\ref{pvol}). This leads to an asymptotic expansion for the energy levels valid 
for large quantum numbers, which should be well-defined. It can be shown, by using the BPS structure of the refined topological string free energy \cite{ikv,ckk}, 
that the quantum B-periods of {\it any} local CY manifold are divergent for an infinite number of values of $\hbar$. In particular, 
the WKB quantization condition for B-periods written down in \cite{acdkv} has an infinite number of poles in the complex $\hbar$-plane. 
 
We conclude that the expression (\ref{pvol}) is {\it incomplete}, and there {\it must} be an 
extra correction which makes the quantum volume of phase space finite and leads to a reasonable quantization condition. 
Since (\ref{ns-p}) already incorporates all the perturbative information available, this correction must be non-perturbative in $\hbar$.

The possibility of having instanton corrections to the quantum volume of phase space was already anticipated in \cite{mp}. 
As pointed out there, in order to understand these corrections we need the geometric approach to the WKB method developed in for 
example \cite{bpv, voros, ddp, dp}. In this approach, the perturbative WKB quantization condition 
is associated to a classical periodic orbit of energy $E$ and the quantum fluctuations around it. The classical action of this trajectory is the classical B-period, 
and the perturbative $\hbar$ corrections 
promote it to a quantum period. In our example, this perturbative analysis leads to the result (\ref{pvol}). 
In fact, the classical periodic orbits can be described in detail. The classical Hamiltonian (\ref{class-H}) leads to the equations of motion
\be
\dot x= {1\over 2} \tanh {p \over 2} , \qquad
\dot p=- {1\over 2}\tanh {x \over 2}. 
\ee
For an orbit of energy $E$, we find 
\be
\dot x= {1\over 2} \sqrt{ 1- 16\,  \re^{-2E} \cosh^2  {x\over 2} }, 
\ee
which can be integrated in terms of Jacobi's elliptic sine, 
\be
\label{real-t}
\tanh  {x \over 2} = k \, \sn \left( {t\over 4}, k\right), 
\ee
where the modulus $k$ is now given by  
\be
k^2= 1- 16\, \re^{-2E}. 
\ee
The function $\sn(u,k)$ is doubly periodic in $u$. It has a real period given by 
\be
\omega_1=4  K(k), 
\ee
where $K(k)$ is the complete elliptic integral of the first kind. The action around this periodic trajectory is 
\be
S_B(E)=  \int_{-8 K(k)}^{8 K(k)} p(t) \dot x(t) \rd t = {\rm vol}_0(E), 
\ee
where we took into account a relative factor of $4$ coming from $u=t/4$. 

\begin{figure}
\center
\includegraphics[height=4cm]{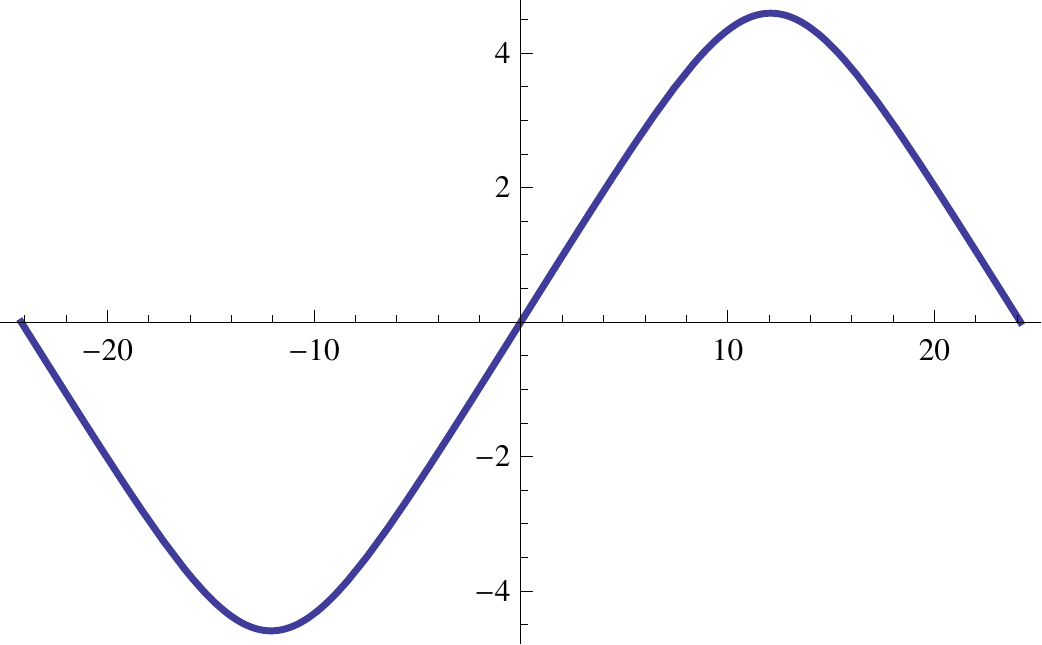}  \qquad \includegraphics[height=4cm]{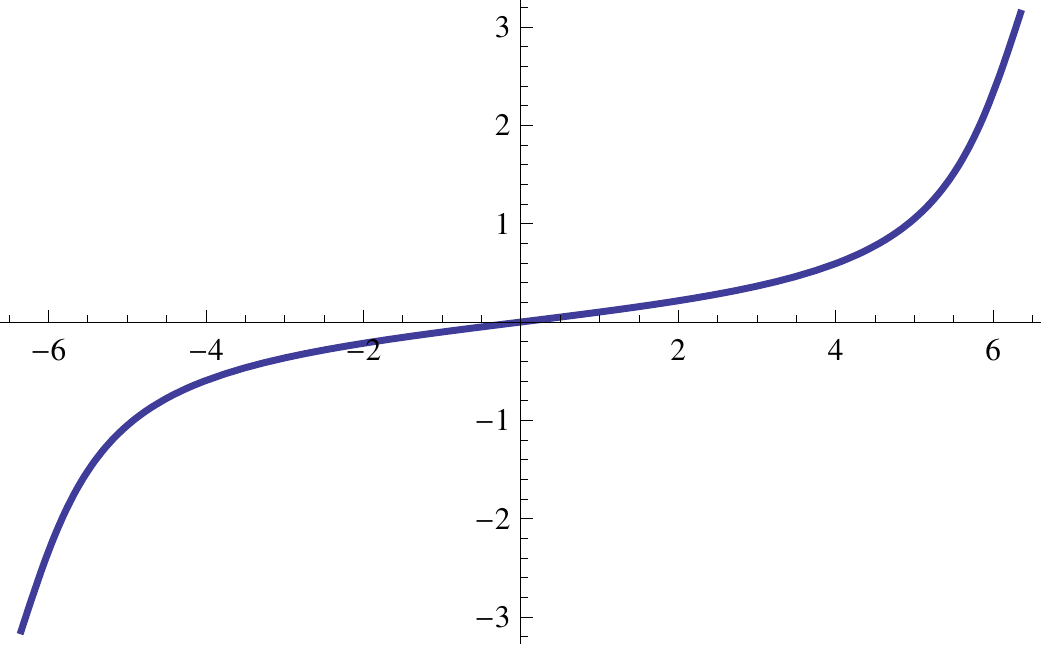} 
\caption{For a given energy $E$, we have a real periodic trajectory given by (\ref{real-t}), which is shown in the figure on the left for $E=3$. 
The horizontal axis represents the time $t$ and it runs through a full period, from $-8 K(k)$ to $8 K(k)$. There is also an imaginary periodic trajectory for $\theta= {\rm Im}(q)$, given by 
(\ref{im-t}), which is shown in the figure on the right, also for $E=3$. Here, the imaginary time $\tau$ runs from $-4 K(k')$ to $4 K(k')$.}
\label{trajs}
\end{figure}

However, as pointed out in \cite{bpv}, in order to obtain a non-perturbative quantization condition one should take into account as well {\it complexified} trajectories. These 
trajectories are allowed because the curve (\ref{class-curve}) has genus one and therefore it has an imaginary period, on top of the real period associated to the real 
periodic orbit (an example of such a situation was discussed recently in \cite{dunne}). In our case this is just the imaginary period of the Jacobi sine function, 
\be
\omega_2=2 \ri K'(k), 
\ee
where
\be 
K'(k)=K(k'), \qquad k'^{2}=1-k^2. 
\ee
In the complexified orbit, time is imaginary, as in \cite{bpv}: $t= \ri \tau$. If we now use the relation 
\be
\sn (\ri u , k)=\ri \,  \tn (u , k'), 
\ee
where $\tn =\sn/ {\rm cn}$, we find that $x$ is also imaginary: $x=\ri \theta$. The equation for the complex trajectory becomes
\be
\label{im-t}
\tan {\theta \over 2}= k \,  \tn \left( {\tau \over 4}, k' \right). 
\ee
The real and complexified trajectories, (\ref{real-t}), (\ref{im-t}), are represented in figure \figref{trajs} for the value of the energy $E=3$. 

\begin{figure}
\center
\includegraphics[height=5.5cm]{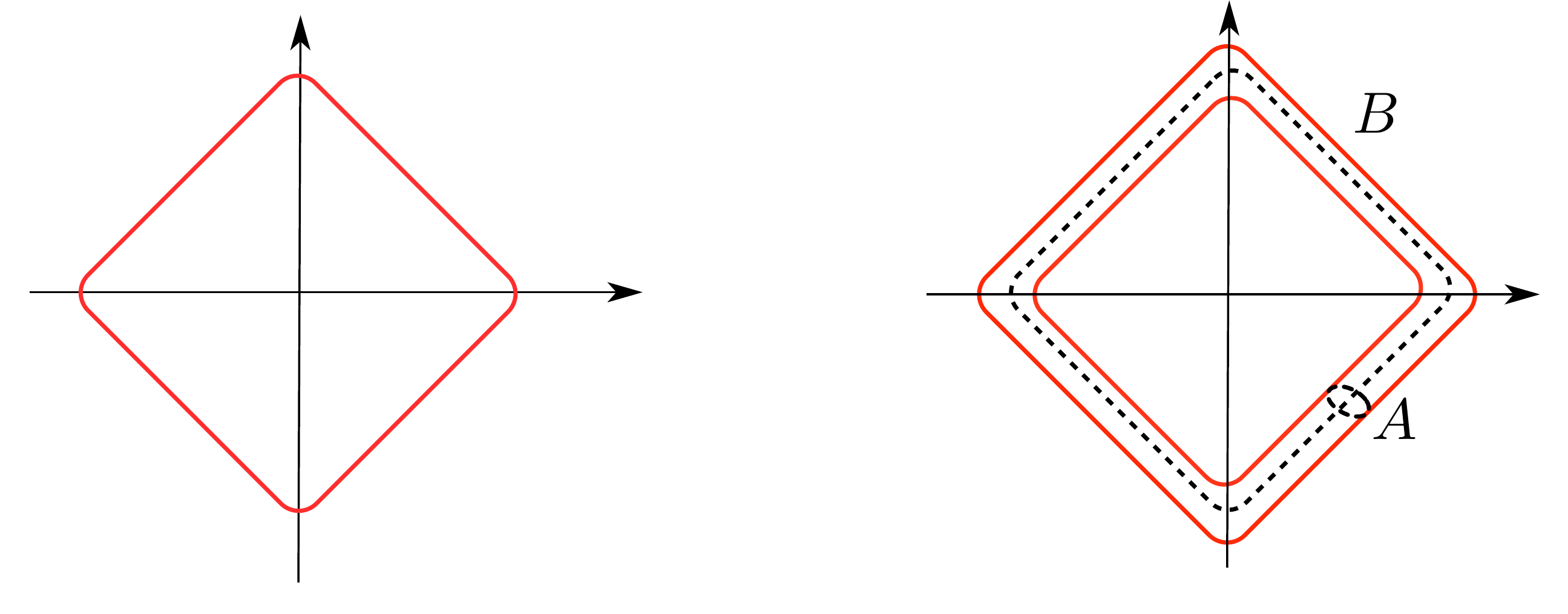} 
\caption{The trajectory (\ref{real-t}) describes a closed orbit in the phase space $(x,p)$, represented schematically in the figure on the left. After complexifying the exponentiated variable $\re^{x/2}$, this closed orbit becomes a torus, as shown in the figure on the right. The imaginary trajectory (\ref{im-t}) is a closed orbit around the $A$-cycle of the torus.}
\label{ab-cycles}
\end{figure}

Geometrically, the trajectory (\ref{real-t}) describes a closed orbit in phase space, along the hypersurface of constant energy $H_{\rm cl}(x,p)=E$. This is the Fermi surface of the ideal Fermi gas 
introduced in \cite{mp}. After complexifying the exponentiated variable $\re^{x/2}$, this closed orbit becomes a torus. The imaginary trajectory (\ref{im-t}) is a closed orbit around the 
$A$-cycle of this torus, while (\ref{real-t}) is now regarded as a 
closed orbit around the $B$-cycle. We depict both orbits in \figref{ab-cycles}. 

In the original Hamiltonian, $\theta$ has a periodicity of $4 \pi$, therefore the relevant action is 
\be 
S_A(E)= 2 \ri  \int_{-4K'(k)} ^{4 K'(k)} p (\tau) \dot \theta (\tau) \rd \tau= -4 \pi \ri \Pi_{A}(z),  
\ee
where we have denoted
\be
\Pi_{A}(z)= \log z + \widetilde \Pi_A (z,z). 
\ee
The contribution of such a complex trajectory to the quantization condition is of the form
\be
\label{sa-period}
\exp \left( {\ri \over \hbar} S_A\right)= \exp \left( { 2 \over k} \Pi_{A}(z) \right). 
\ee
Including the quantum corrections simply promotes the classical period to its quantum counterpart, as already noticed in \cite{bpv}. As in \cite{ddp,dp}, we will call 
the exponentiated quantum period associated to a cycle a {\it Voros multiplier}. The quantum A-period, specialized to the ``slice" (\ref{zz-q}), is 
\be
\Pi_{A_I}(q^{1/2} z, q^{-1/2} z; \hbar)= \log z \pm {\ri \pi k \over 2} + \sum_{\ell \ge 1} \widehat a_\ell(\hbar) z^\ell, \qquad I=1,2, 
\ee
where the $\pm$ sign corresponds to $I=1,2$, respectively. We conclude from (\ref{sa-period}) that the appropriate Voros multiplier for the A-period in this theory is 
\be
\label{A-vm}
 \exp \left[ {2\over k} \Pi_{A_I}(q^{1/2} z, q^{-1/2} z; \hbar)\right]= - \re^{-4 E_{\rm eff}/k}, 
\ee
where
\be
\label{Effi}
E_{\rm eff}=   E -{1 \over 2} \sum_{\ell=1}^\infty \widehat a_\ell(\hbar) \re^{-2 \ell E}.
\ee
This result was also obtained in \cite{mp}\footnote{In \cite{mp}, what is here called the A-period was called the B-period, and viceversa. The notation we are using agrees with the standard conventions for periods in local Calabi--Yau manifolds.}. In general, the non-perturbative correction to the quantum volume is a formal power series in the Voros multiplier for the quantum A-period. In our case, it takes the form, 
\be
\label{sm-series}
{\rm vol}_{\rm np} (E; \hbar)= \sum_{m=1}^\infty s_m (k) (-1)^m  \re^{-4 m  E_{\rm eff}/k}.
\ee
This is clearly non-perturbative in $\hbar$ (or equivalently, in $k$), and it is invisible in the standard perturbative correction to the WKB condition. 

The calculation of $s_m (k)$ from first principles is difficult. In the case of the Schr\"odinger equation studied in \cite{bpv, voros, ddp, dp}, 
the instanton corrections are determined by the vanishing of the so-called Jost function, and their calculation 
requires a detailed analysis of the spectral problem and of the WKB wavefunction. We will now present a conjecture for the form of the instanton corrections for the spectral problem (\ref{diff-eq}).  
To write down our formula, let us consider the worldsheet instanton corrections to the un-refined 
topological string free energy on local $\IP^1 \times \IP^1$, in the Gopakumar--Vafa form \cite{gv}: 
\be
\label{gv-exp}
F(T_1, T_2, g_s)=\sum_{g\ge0}\sum_{w\ge 1} \sum_{d_1, d_2} 
{(-1)^{g-1} \over w} n _g^{d_1, d_2} \left( q_s^{w/2} -q_s^{-w/2} \right)^{2g-2} {\rm e}^{-w (d_1 T_1 + d_2 T_2)}.
\ee
Here, 
\be
q_s= \re^{g_s},
\ee
and $g_s$ is the topological string coupling constant. In (\ref{gv-exp}), $T_1$, $T_2$ are the complexified K\"ahler classes, corresponding to the two compact 
$\IP^1$s in the geometry, and $n_g^{d_1, d_2}$ are the Gopakumar--Vafa invariants of local $\IP^1 \times \IP^1$ for genus $g$ and degrees $d_1, d_2$. Along the ``slice" 
\be
\label{gv-slice}
T_1=T_2=T
\ee
 the natural invariant is the diagonal one, 
\be
\label{gvtwo}
n^d_g= \sum_{d_1+d_2 =d} n^{d_1, d_2}_g. 
\ee

We are now ready to state our conjecture about the form of the instanton corrections. We claim that the coefficients $s_m(k)$ in (\ref{sm-series}) are given by
\be
\label{sm}
s_m(k)= -4 \pi k\sin \left({4 \pi m \over k} \right) d_m(k), 
\ee
where 
\be
\label{gvone}
d_m(k)= {1   \over m} \sum_{g\ge 0} \sum_{d|m} d \, n^d_g  \left( 2 \sin {2 \pi m \over d k} \right)^{2g-2} .
\ee
The coefficients $d_m(k)$ have a simple interpretation: they are the coefficients of $\re^{-mT}$ in the topological string free energy of local $\IP^1 \times \IP^1$, along the slice (\ref{gv-slice}), and for
\be
g_s= {4 \pi \ri \over k}. 
\ee
We then conjecture that the non-perturbative contribution to the quantum volume of phase space is 
\be
\label{np-vol}
{\rm vol}_{\rm np} (E; \hbar)=-4 \pi k  \sum_{m=1}^\infty \sin \left({4 \pi m \over k} \right) (-1)^m d_m(k) \re^{-4 m  E_{\rm eff}/k}, 
\ee
and the total quantum volume is 
\be
\label{total-vol}
{\rm vol}(E;\hbar)= {\rm vol}_{\rm p} (E; \hbar)+ {\rm vol}_{\rm np} (E; \hbar). 
\ee
Notice that the non-perturbative volume (\ref{np-vol}) is also divergent for integer values of $k$. We will now show that the total volume (\ref{total-vol}) is well-defined, 
i.e. the divergences in ${\rm vol}_{\rm p}(E;\hbar)$ cancel against the divergences in ${\rm vol}_{\rm np} (E; \hbar)$. 

Before showing this, let us give some indications on the origin of this conjecture. As we will see in the next section, the spectral problem (\ref{diff-eq}) appears in the Fermi gas approach to ABJM theory. It turns out that the grand potential of ABJM theory is closely related to the volume of phase space. The perturbative part (\ref{pvol}) leads to the non-perturbative membrane corrections to the grand potential, while the non-perturbative part (\ref{np-vol}) leads to the worldsheet-instanton corrections to the grand potential. The conjecture (\ref{np-vol}) is inspired by the known form of these corrections \cite{mpabjm, dmp,hmo1}. The requirement that divergences should cancel in the total volume (\ref{total-vol}) is a dual manifestation of the HMO cancellation mechanism discovered in \cite{hmo1}. According to this mechanism, the divergences in the worldsheet instanton part of the grand potential of ABJM theory should cancel against the divergences in the membrane instanton part, since the total grand potential is well-defined and finite for any $k$. We have here a similar mechanism, which is based this time on the fact that the spectral problem is well-defined for any value of $k$. The cancellation mechanism in the quantum volume is simpler however than the HMO mechanism, since it only involves simple poles, while the HMO mechanism involves double poles. 

Let us now verify that ${\rm vol}(E; \hbar)$ is well-defined for any value of $k$. Since the non-perturbative contribution is defined in terms of $E_{\rm eff}$, instead of $E$, let us re-express the perturbative part ${\rm vol}_{\rm p}(E; \hbar)$ in terms of this variable. This defines a new set of coefficients ${\widetilde b}_\ell (k)$ as follows, 
\be
\label{pvolb}
{\rm vol}_{\rm p}(E; \hbar)= 8 E_{\rm eff}^2 -{4 \pi^2 \over 3} +{\hbar^2 \over 24}  + 4 \pi^2 k  \sum_{\ell=1}^\infty {\widetilde b}_\ell (k) \re^{-2 \ell E_{\rm eff}}.
\ee
These coefficients were first introduced in \cite{hmo2}, in the context of ABJM theory, and their geometric meaning was uncovered in \cite{hmmo}: 
they can be expressed in terms of the refined BPS invariants $N_{j_L, j_R}^{d_1, d_2}$ of local $\IP^1 \times \IP^1$ as
\be
\label{blj}
\widetilde{b}_\ell(k)=-\frac{\ell}{2\pi}\sum_{j_L,j_R}\sum_{\ell=dw}\sum_{d_1+d_2=d}N^{d_1,d_2}_{j_L,j_R}q^{\frac{w}{2}(d_1-d_2)}
\frac{\sin\frac{\pi kw}{2}(2j_L+1)\sin\frac{\pi kw}{2}(2j_R+1)}{w^2\sin^3\frac{\pi kw}{2}}. 
\ee
This formula is based on the fact that the combination of B-periods appearing in (\ref{pvol}) is a derivative of the refined topological string free energy, which in turn can be written in terms of 
refined BPS invariants (see for example \cite{ckk} for a summary of these facts, and a list of values of the refined invariants for low degrees). On the other hand, 
the coefficient (\ref{gvone}) can be expressed 
in terms of the same invariants by the formula, 
\be
 d_m(k)=\sum_{j_L,j_R}\sum_{m=dn}\sum_{d_1+d_2=d}N^{d_1,d_2}_{j_L,j_R}
\frac{2j_R+1}{(2\sin\frac{2\pi n}{k})^2}\frac{\sin\left( \frac{4\pi n}{k}(2j_L+1) \right)}{\sin\frac{4\pi n}{k}}\frac{1}{n}, 
\ee
see \cite{hmmo} for a derivation. 
We can now use a simplified version of the argument appearing in \cite{hmmo} to check that the singularities in (\ref{total-vol}) cancel. 
First of all, notice that the singularities appear when $k$ takes 
the form 
\be
\label{ksing}
k={2n \over w}={2m \over \ell}.
\ee
 The singularities are simple poles. The poles appearing in ${\rm vol}_{\rm np}(E; \hbar)$ are of the form 
 \be
 \label{ws-pole}
  (-1)^m  {8 m \over   w^3 \left(  k -{2n \over w} \right)} (1+2 j_L)(1+ 2j_R) 
N^{d_1,d_2}_{j_L,j_R} \re^{-{2 m w \over n}  E_{\rm eff} }.
\ee
The corresponding poles appearing in (\ref{pvolb}) are of the form
\be
\label{b-pole}
-  \re^{\pi \ri k w(d_1-d_2)/2} {8 m \over w^3 \left( k -{2n \over w} \right)} (-1)^{n(2j_L+2j_R-1)} (1+2 j_L)(1+ 2j_R)  N^{d_1,d_2}_{j_L,j_R}\re^{-2\ell E_{\rm eff}}.
\ee
By using (\ref{ksing}), one notices that 
\be
 \re^{\pi \ri k w(d_1-d_2)/2}=(-1)^m,
\ee
and it is easy to see that all poles in (\ref{ws-pole}) cancel against the poles in (\ref{b-pole}), for any value of $E_{\rm eff}$, provided that
\be
(-1)^{n(2j_L+2j_R-1)}=1.
\ee
This can be seen to be the case by a geometric argument explained in \cite{hmmo}. We conclude that ${\rm vol}(E; \hbar)$ is well-defined and finite for any real value of $\hbar$, as a series in $\re^{-2E}$.  

The exact WKB quantization condition reads now
\be \label{volbs}
{\rm vol}(E; \hbar)= 2 \pi \hbar \left( n +{1\over 2} \right),\qquad n=0, 1,2, \cdots,
\ee
where ${\rm vol}(E; \hbar)$ is a sum of the perturbative part $\eqref{pvol}$ and the non-perturbative part $\eqref{np-vol}$. This condition determines the energy levels $E_n$ as functions of $n$ and $\hbar$. As in similar examples of exact WKB quantization conditions, the total ${\rm vol}(E)$ is a trans-series 
involving various small parameters, on top of $\hbar$ itself. 
On one hand we have of course $\hbar$, 
but we also have exponentially small quantities in $\hbar$, and non-analytic functions of $\hbar$ at $\hbar=0$, like the trigonometric functions of $1/k$ appearing in (\ref{np-vol}). We can solve this quantization quantization at small $\hbar$, as one does for example in the case of the double-well potential in Quantum Mechanics \cite{zjj}. 
To give a flavour of the type of expressions one finds, let us calculate the first non-perturbative correction to the perturbative series in (\ref{enseries}). In order to 
do this calculation, we need the leading term of $E_{\rm eff}$ in an expansion around $\hbar=0$. We find, after using (\ref{aobo}), 
\be
E_{\rm eff}(E)= E_{\rm eff}^{(0)}(E)+ \CO(\hbar^2), 
\ee
where
\be
E_{\rm eff}^{(0)}(E)= E-2 \re^{-2E} \, _4F_3\left(1,1,\frac{3}{2},\frac{3}{2};2,2,2;16  \re^{-2E}\right). 
\ee
After expanding it around $E=2 \log 2$, we find,  
\be
E_{\rm eff}^{(0)}(E)= {4 {\rm K} \over \pi} + {1\over \pi} \left(E -2 \log 2\right) \left(1 - \log \left( {E -2\log 2 \over 8} \right)  \right)+\CO\left( \left(E -2 \log 2\right)^2\right),
\ee
where ${\rm K}$ is the Catalan number. A simple calculation shows that 
\be
E_n = E_n^{\rm p} + E_n ^{\rm np}, 
\ee
where, as $\hbar \rightarrow 0$, 
\be
E_n^{\rm np} \approx -{1\over 4 \pi} \left( {2n+1 \over 64}\right)^{2n+1} \sin \left( {4 \pi \over k}\right) d_1(k) \re^{-2n-1} \hbar^{2n+2} \re^{-A/k}, 
\ee
and 
\be
 A=  {16 {\rm K} \over \pi}. 
 \ee
It seems that $E_n ^{\rm np}$ is a trans-series involving the small parameter $\re^{-A/k}$, $\log \hbar$ and trigonometric functions of $1/k$.

Before closing this subsection, let us summarize the two most important consequences of our proposal for the exact quantization condition: 

\begin{enumerate} 

\item In solving the spectral problem (\ref{diff-eq}) associated to the quantum curve of local $\IP^1 \times \IP^1$, the information encoded in the quantum periods (or, equivalently, in the NS limit of the refined topological string), evaluated with the perturbative WKB method as in \cite{mm,acdkv}, is {\it not} enough. The perturbative quantization condition is insufficient and even ill-defined for some values of $\hbar$, and has to be corrected by an infinite series of instantons. In the terminology of \cite{ddp, dp}, we can say that the WKB method of \cite{mm,acdkv} gives the Voros multipliers of the problem, but does not contain the information about the Jost function solving the spectral problem. 

\item The information about the instanton corrections involves the {\it standard} topological string free energy, and in particular the standard Gopakumar--Vafa invariants. 
This is a statement dual to the conjecture in \cite{hmmo}, where evidence was given that the non-perturbative corrections to the standard topological string free energy involves the refined topological string. The fact that the un-refined topological string and the NS refined topological string are intimately related in a non-perturbative treatment is reminiscent of the observations made in \cite{lv}. 
\end{enumerate}

Although our conjecture is well-motivated by the relationship to ABJM theory, it can be tested in detail. After all, our conjecture gives the exact WKB quantization condition of the spectral problem, 
and we can (and should) compare it to the actual values of the energy levels. 
 
\subsection{Testing the non-perturbative WKB quantization condition}

The goal of this subsection is to compute the energy levels using (\ref{volbs}), but this time at large quantum number $n$ and fixed $k$. 
In this computation the non-perturbative part of ${\rm vol}(E;\hbar)$ plays a crucial role. We will then compare the energy levels obtained in this way with numerical values 
obtained directly from the integral equation $\eqref{int-eq}$, and find an impressive agreement.

Since the numerical spectrum is easier to compute with high accuracy when $k=1,2$, we will study (\ref{volbs}) for these values of $k$ (corresponding to $\hbar=2\pi$ and $\hbar=4\pi$, respectively). 
In these cases, we can see from $\eqref{pvol}$ and $\eqref{np-vol}$ that ${\rm vol}(E; \hbar)$ is given as an expansion in powers of $\re^{-2E}$. In fact, for $k=1$, the coefficients of the odd powers 
\be
\re^{-2(2m +1) E}~, \quad  m=0,1,2,\ldots, 
\ee
vanish. Therefore, for $k=1$ we will have an expansion in powers of $\re^{-4E}$. Let us write the total quantum volume, for $k=1$ and $k=2$, as 
\be
\label{vol-exact}
{ {\rm vol}(E; \hbar) \over 2 \pi \hbar} = \widehat{C}(\hbar) E^2 + \widehat{n}_0(\hbar) + \sum_{\ell=1}^\infty \left( A_\ell (\hbar) E + B_\ell (\hbar) \right) \re^{- \frac{4 \ell E}{k}},
\ee
where
\be
\widehat{C}(\hbar)=\frac{4}{\pi\hbar}~, \qquad 
\widehat{n}_0(\hbar)=-\frac{2\pi}{3\hbar}+\frac{\hbar}{48\pi}~.
\ee
The coefficients $B_\ell(\hbar)$ will in general have contributions from both the perturbative and the non-perturbative part.

Since the volume is given as a large $E$ expansion, we will calculate the energy levels in an asymptotic expansion for large quantum numbers. Let us assume an ansatz for the solution of the exact 
WKB quantization condition of the  form 
\be \label{enasymp}
E_n = E_n ^{(0)}+ \sum_{\ell=1}^\infty E_n^{(\ell)}\re^{-\frac{4 \ell E_n^{(0)}}{k}}.
\ee
Plugging it into $\eqref{volbs}$ we find, 
\be
\ba
E_n^{(0)}&=\sqrt{\frac{n+1/2-\widehat{n}_0(\hbar)}{\widehat{C}(\hbar)}}, \\
E_n^{(1)}&=-\frac{1}{2\widehat{C}(\hbar)E_n^{(0)}}\left(B_1(\hbar)+A_1(\hbar) E_n^{(0)}\right), \\
E_n^{(2)}&=\frac{1}{2\widehat{C}(\hbar)E_n^{(0)}}\Bigg( -B_2(\hbar)-A_2(\hbar)E_n^{(0)}  + \frac{4}{k}A_1(\hbar) E_n^{(1)}E_n^{(0)}-\widehat{C}(\hbar) (E_n^{(1)})^2 \\
&-A_1(\hbar)E_{n}^{(1)}+\frac{4}{k}B_1(\hbar) E_n^{(1)}\Bigg), 
\ea
\ee
as well as
\be
\ba
E_n^{(3)}&=\frac{1}{2 \widehat{C}(\hbar) E_n^{(0)}}\Bigg(-B_3(\hbar)-A_3(\hbar) E_n^{(0)}-A_2(\hbar)\left(E_n^{(1)} -\frac{8}{k} E_n^{(0)} E_n^{(1)}\right)\\
&+\frac{8}{k} B_2(\hbar) E_n^{(1)} -2 \widehat{C}(\hbar)E_n^{(1)}E_n^{(2)} +B_1(\hbar)\left(\frac{4}{k}E_n^{(2)}-\frac{8}{k^2}(E_n^{(1)})^2\right)\\
&+A_1(\hbar)\left( \frac{4}{k} (E_n^{(1)})^2-\frac{8}{k^2}E_n^{(0)}(E_n^{(1)})^2-E_n^{(2)}+\frac{4}{k}E_n^{(0)} E_n^{(2)}\right) \Bigg)~.
\ea
\ee
To lowest order, plugging in the values we find
\be
E_n^{(0)}={ \sqrt{k}\pi \over {\sqrt{2}}} \left( n + {1\over 2} +{1\over 3k } - {k \over 24} \right)^{1/2}. 
\ee
This expression is valid for any $k>0$, provided $n$ is large enough. The leading growth of the eigenvalues derived from the lowest order approximation, 
\be
E^2_n- E_0^2 \approx {k\pi^2 \over 2} n, \qquad n\gg 1, 
\ee
has been verified numerically in \cite{hmo} for $k=1$, by computing the spectrum for $n=0, \cdots, 6$. 

In order to obtain subleading corrections we have to calculate the coefficients $A_\ell(\hbar)$ and $B_\ell(\hbar)$, which are determined by the coefficients 
$\widehat{a}_\ell(\hbar)$, $\widehat{b}_\ell(\hbar)$ in (\ref{pvol}) 
and $s_m (k)$ in (\ref{np-vol}). For the first few values of $\ell,m$, they can be read off from $\eqref{q-ABper}$ and $\eqref{q-ABperrelab}$, and from 
the Gopakumar--Vafa invariants of local $\IP^1 \times \IP^1$, as listed in for example \cite{hmo1}. For the first three orders we find
\be
\begin{split}
\frac{{\rm vol}(E;\hbar=2\pi)}{4\pi^2}&=\frac{2}{\pi^2}E^2-\frac{7}{24}+\frac{8}{\pi^2}E\re^{-4E}+\frac{1}{\pi^2}\re^{-4E}-\frac{52}{\pi^2}E \re^{-8E} -\frac{1}{4\pi^2}\re^{-8E} \\
&+\frac{1472}{3\pi^2}E \re^{-12E}-\frac{152}{9\pi^2}\re^{-12E}+\mathcal{O}(E\re^{-16E})~, \\
\frac{{\rm vol}(E;\hbar=4\pi)}{8\pi^2}&=\frac{1}{\pi^2}E^2-\frac{1}{12}+\frac{8}{\pi^2}E\re^{-2E}+\frac{2}{\pi^2}\re^{-2E}-\frac{52}{\pi^2}E \re^{-4E} -\frac{1}{2\pi^2}\re^{-4E} \\
&+\frac{1472}{3\pi^2}E \re^{-6E}-\frac{304}{9\pi^2}\re^{-6E}+\mathcal{O}(E\re^{-8E}) ~,
\end{split}
\ee
and the poles cancel, as expected.

Explicit calculations of the coefficients $\eqref{q-ABperrelab}$ and (\ref{gvone}) in \cite{hmmo} indicate that (\ref{vol-exact}) 
is a convergent series when $|q|=1$ and for sufficiently large $E$. Notice that, for small $\hbar$, i.e. $q\rightarrow 1$, the coefficients in $\eqref{q-ABperrelab}$ are just 
the coefficients of the classical B-period, which are known to lead to a series with a finite radius of convergence. It seems that these convergence properties are 
preserved as long as $q$ stays in the unit circle of the complex plane. It follows that the series (\ref{enasymp}) should also be convergent, for $n$ large enough. 
Our numerical calculations indicate that, in fact, this series converges very rapidly already for $n=0, 1$. The results for $E_n$, for $n=0,1$, as computed with the WKB quantization condition, and up to third order for $k=1$ and $k=2$ are listed in Tables \ref{k1table} and \ref{k2table}, respectively. 

\begin{table}[t] 
\centering
\begin{tabular}{l l l}
&Energy levels for $k=1$&\\
\hline
Order &	$E_0$ & $E_1$ \\
\hline
$0$ & ${\underline{1.97}}654203314$ & \underline{2.9734}69456\\
$1$ & ${\underline{1.97575}}850097$ & \underline{2.97345521}8 \\
$2$ & ${\underline{1.975757951}}36$ &\underline{2.973455217}\\
$3$ & ${\underline{1.97575795105}}$ & \underline{2.973455217}\\
\hline
Numerical value & 1.97575795105 & 2.973455217
\end{tabular}
\caption{The lowest and next-to-lowest energy eigenvalues for $k=1$ calculated analytically, including higher and higher orders of exponentially small corrections in $\eqref{enasymp}$. In the last line numerical values are given. At each order of the approximation, we underline the digits which agree with the numerical result. }
 \label{k1table}
\end{table}%
\begin{table}[t] 
\centering
\begin{tabular}{l l l}
&Energy levels for $k=2$&\\
\hline
Order &	$E_0$ & $E_1$ \\
\hline
$0$ & $\underline{2.3}99431022965$ & \underline{3.95}3084066277\\
$1$ & $\underline{2.36}3040773485$ & \underline{3.95151}7001949 \\
$2$ & $\underline{2.3623}88178770$ & \underline{3.951515902}713\\
$3$ & $\underline{2.362377}640277$ & \underline{3.951515902099}\\
\hline
Numerical value & 2.362377493014 &3.951515902099
\end{tabular}
\caption{The lowest and next-to-lowest energy eigenvalues for $k=2$ calculated analytically, including higher and higher orders of exponentially small corrections in $\eqref{enasymp}$. In the last line numerical values are given. At each order of the approximation, we underline the digits which agree with the numerical result. }
 \label{k2table}
\end{table}%

We can now compare these results with numerical values obtained starting from the integral equation $\eqref{int-eq}$. In order to get numerical values with high accuracy we will use that, as shown in \cite{hmo}, this equation can be rewritten in the form of an eigenvalue equation for an infinite dimensional matrix $M$. For the derivation we refer to \cite{hmo}, here we only quote the results that we need in order to numerically test the energy eigenvalues we have obtained. The matrix elements $M_{nm}$ depend only on $m+n$. Such a matrix is called a Hankel matrix. Moreover, in this case they are zero if $m+n$ is odd, that is, $M$ has the following form
\be
M=\begin{pmatrix}
m_0 & 0 & m_1 & 0 & m_2 & 0 & \ldots \\
0 & m_1 & 0 &  m_2 &0 & m_3  & \\
m_1 & 0 &  m_2 &0 & m_3  & 0 &  \\
0 &  m_2 &0 & m_3  & 0 & m_4 & \\
m_2 & 0 & m_3  & 0 & m_4 & 0 & \\
\vdots &  & & &  &  &\ddots
\end{pmatrix}~.
\ee
Such a matrix can be decomposed into two blocks of Hankel matrices, $M_+$ and $M_-$:
\begin{align}
M_+= \begin{pmatrix}
m_0 & m_1 & m_2 & \ldots \\
m_1 &  m_2 & m_3  & \\
m_2 & m_3 & m_4  \\
\vdots &    &  &\ddots
\end{pmatrix}, \qquad  & M_-= \begin{pmatrix}
m_1 & m_2 & m_3 & \ldots \\
m_2 &  m_3 & m_4  & \\
m_3 & m_4 & m_5  \\
\vdots &    &  &\ddots
\end{pmatrix}~.
\end{align}
The eigenspaces of $M$ decompose into a direct product of the eigenspaces of $M_\pm$. Let the eigenvalues of $M$ be denoted by $\lambda_n$, ordered such that
\be
\lambda_0 >\lambda_1 >\lambda_2 >\ldots~,
\ee 
and let the eigenvalues of $M_\pm$ be denoted by $\lambda_{\pm,n}$, ordered in the same way. We then have
\begin{align}
\lambda_{+,n}=\lambda_{2n}, \quad \lambda_{-,n}=\lambda_{2n+1}~.
\end{align}
The relation between the eigenvalues of $M$ and the energy eigenvalues is
\be
E_n=-\log{\lambda}_n~.
\ee
Different values of $k$ give different $M_{nm}$. For $k=1$ we have (for $m+n$ even, otherwise we get zero)
\be
M^{k=1}_{nm}=\frac{C_{\frac{m+n}{2}}}{2^{n+m+3}}
 \label{Mk1}
\ee 
where $C_{n}$ is the Catalan number
\be
C_n=\frac{(2n)!}{(n+1)!n!}~.
\ee
For $k=2$ we have (again for $m+n$ even, otherwise we get zero)\footnote{Note that there is a factor of 2 missing in equation 2.41 in the first version of \cite{hmo}.}
\be
M^{k=2}_{nm}=\frac{1}{4\pi}\left[-\frac{2}{n+m+1}+\psi\left(\frac{n+m+3}{4}\right)-\psi\left(\frac{n+m+1}{4}\right)\right] \label{Mk2}
\ee
where $\psi(x)$ is the digamma function.

We have calculated the lowest and next-to-lowest energy eigenvalues obtained from these two matrices numerically. In principle, we have to diagonalize an infinite-dimensional matrix, but in practice we have to truncate the Hankel matrices to an $L\times L$ matrix. The eigenvalues of this truncated matrix $E_n (L)$ give numerical approximations to the exact $E_n$, and they converge to it as $L \rightarrow \infty$. In order to incorporate finite-size effects, let us assume that the eigenvalues depend on $L$ as
\be
E_{n}(L) =E_n + \sum_{j\ge 1} {E_n^j  \over L^j}. 
\ee
We can accelerate the convergence of the sequence $E_{n,L}$ to $E_n$ by using for example Richardson extrapolation \cite{bo}. The results 
obtained with this method for $k=1$ and $k=2$ are given in Table \ref{k1table} and \ref{k2table}, respectively, with the 
displayed numerical accuracy.  In all cases, the series of instanton 
corrections add up to values closer and closer to the numerical value. 
For $k=1$, and for the first excited state with $k=2$, the exponential corrections of fourth order and higher are not visible in the numerical approximation 
to the eigenvalue, due to the limited accuracy of our calculation. In the case of the ground state energy $E_0$ for $k=2$, 
the corrections are moderately large: the approximation obtained by including up to 
the third exponentially small correction is only correct to order $10^{-7}$. 
We have checked that including the fourth order correction gives the correct numerical eigenvalue to order $10^{-9}$.

It seems clear from this numerical analysis that our ansatz (\ref{np-vol}) for 
the non-perturbative corrections to the quantum volume is not only divergence-free: it is also in excellent agreement 
with the true spectrum of eigenvalues.

\sectiono{The grand potential of ABJM theory}

As we explained in the introduction, the second motivation for looking at the spectral problem (\ref{int-eq}) 
is the study of non-perturbative effects in the partition function 
of ABJM theory \cite{abjm}. This partition function $Z(N,k)$ depends on two parameters: the rank of the gauge group $N$ 
and the Chern--Simons level $k$. Building on \cite{kwy,kwy-2}, it was shown in \cite{mp} that $Z(N,k)$ can be written as 
\be
\label{zabjm}
Z(N,k)={1 \over N!} \sum_{\sigma  \in S_N} (-1)^{\epsilon(\sigma)}  \int  \rd ^N x \prod_i \rho(x_i, x_{\sigma(i)}), 
\ee
where $\rho(x_1, x_2)$ is the kernel defined in (\ref{rhok}). This is nothing but the partition function of an ideal Fermi 
gas with one-particle energies $E_n$, $n\ge 0$, determined by the spectral problem (\ref{int-eq}). The grand potential of the Fermi gas is given by 
\be
J(\mu, k)= \sum_{n \ge 0} \log \left(1 + \re^{\mu-E_n} \right), 
\ee
where the energy levels $E_n$ are determined by the WKB quantization condition (\ref{volbs}), which defines in fact an implicit function $E(n)$ for arbitrary 
values of $n$. In order to perform the sum over discrete energy levels, we will use 
the Euler--Maclaurin formula, which reads
\be
\label{emc}
\sum_{n\geq 0}f(n)=\int_{0}^\infty f(n) \rd n+\frac{1}{2}\left(f(0)+f(\infty)\right)+\sum_{r\geq 1} \frac{B_{2r}}{(2r)!}\left(f^{(2r-1)}(\infty)-f^{(2r-1)}(0)\right). 
\ee
Notice that, in general, this formula gives an asymptotic expansion for the sum. However, as noticed in \cite{cg}, this asymptotic expansion is Borel summable under mild assumptions 
for the function $f(n)$, and this can be used to write an exact formula. In this paper we will not explore this possibility. In our case, the function $f(n)$ is given by 
\be
f(n)= \log{\left(1+ \re^{\mu-E(n)}\right)}. 
\ee
Since $E(n)\rightarrow \infty$ as $n\rightarrow\infty$, we have $f(\infty)=0$, $f^{(2r-1)}(\infty)=0$ for all $r \ge 1$. The first terms of (\ref{emc}) give, 
\be
\int_{E_0}^\infty \frac{\rd n(E)}{\rd E}\log{\left(1+\re^{\mu-E}\right)}\rd E+\frac{1}{2}f(0)=\frac{1}{2\pi\hbar}\int_{E_0}^\infty \frac{ {\rm vol}(E)}{1+\re^{E-\mu}}\rd E. 
\ee
In deriving this equation, we first changed variables from $n$ to $E$, used 
\be
n(E)=\frac{{ \rm vol}(E)}{2\pi\hbar}-\frac{1}{2},
\ee
we integrated by parts, and we took into account that 
\be
\frac{{\rm vol}(E_0)}{2\pi\hbar}=\frac{1}{2}, 
\ee
as well as the asymptotic behavior 
\be
{\rm vol}(E)\approx E^2, \qquad E\rightarrow \infty.
\ee
We conclude that\footnote{In the first version of this paper, we didn't include the corrections involving the derivatives of $f(n)$ at $n=0$, and some of the resulting 
formulae were incorrect, as pointed out to us by Yasuyuki Hatsuda. We would like to thank him for his precious observations, which prompted us to find the correct formula for the grand potential.}
\be
\label{j-int}
J(\mu, k)={1\over 2 \pi \hbar}  \int_{E_0}^\infty {{\rm vol}(E) \rd E \over \re^{E-\mu}+1}-\sum_{r\geq 1} \frac{B_{2r}}{(2r)!}f^{(2r-1)}(0).
\ee
The first term in this formula is nothing but an integral transform of the quantum volume 
of phase space studied in the previous section. In order to analyze it, we just have to compute the integrals 
\be
\label{rjm}
R_\ell ^{(j)}(E_0, \mu)=\int_{E_0} ^\infty {E^j \re^{- 2 \ell E} \over \re^{E-\mu} +1}  \rd E, \qquad j=0,1, \qquad  \ell \in \IN, 
\ee
which appear when considering the perturbative quantum volume (\ref{pvol}), as well as the integrals 
\be
\label{r-sig}
R_{\sigma}(E_0, \mu)= \int_{E_0} ^\infty { \re^{- \sigma E} \over \re^{E-\mu} +1} \rd E,
\ee
where $\sigma \notin \IN$. These appear when we consider the non-perturbative quantum volume (\ref{np-vol}) for $k$ arbitrary. 

The calculation of these integrals is not elementary, and we will use a variant of the Mellin transform used in \cite{zjj}. It is defined as 
\be 
\label{mt}
\widehat g(s)= \int_0^1 g(u) u^{-s-1} \rd u. 
\ee
This will make it possible to calculate analytically the integral over the density of states.

We will first consider the contribution to the integral due to the perturbative quantum volume. In order to have a more compact notation for the answer, it is useful to define the functions 
\be
\ba
\widehat a_\ell (s; \hbar)&= \widehat a_\ell (\hbar) \re^{-(2\ell-s) E_0} \left (1+ \left( 2\ell-s \right) E_0\right), \\
\widehat b_\ell (s; \hbar)&= \widehat b_\ell (\hbar) \re^{-(2\ell-s) E_0},
\ea
\ee
as well as the elementary integrals 
\be
\label{cijn}
\CI_j(n)= \int_{E_0}^\infty E^j \re^{-n E} \rd E = \left( -{\partial \over \partial n} \right)^j  \left( {1\over n} \re^{-n E_0} \right), \qquad j \ge 0.
\ee
Using the results collected in the Appendix, one finds 
\be
\label{jp-mu}
\ba
 {1\over 2 \pi \hbar}  \int_{E_0}^\infty {{\rm vol}_{\rm p} (E) \rd E \over \re^{E-\mu}+1} &= {2 \mu^3 \over 3 k \pi^2} + \left( {1\over 3k} + {k \over 24}\right) \mu+ \widehat A(\hbar) \\
&\, \, + \sum_{\ell \ge 1} \left( -{\widehat a_\ell (\hbar) \over \pi^2 k} \mu^2  + {\widehat b_\ell (\hbar) \over 2 \pi^2 k} \mu  + { \widehat c_\ell (\hbar) 
\over 2 \pi^2 k}  \right)\re^{-2 \ell  \mu}- \sum_{\ell \ge 0}  {  \widehat{d}_\ell (\hbar) 
\over 2 \pi^2 k} \re^{-(2 \ell+1) \mu}. 
\ea
\ee
In this expression, the $\mu$-independent function $\widehat{A}(\hbar)$ is given by 
\be
\label{ak}
\widehat A(\hbar)= -{1\over 2 \pi^2 k } \left[ {4 \over 3} E_0^3 - \left({2 \pi^2 \over 3}-{\hbar^2 \over 48} \right) E_0 +  \sum_{\ell \ge 1} {\widehat a_\ell (0; \hbar) \over  \ell^2} - \sum_{\ell \ge 1} {\widehat b_\ell (0; \hbar) \over 2 \ell} \right],
\ee
and the coefficients $\widehat c_\ell$ and $\widehat{d}_\ell$ are given by 
\be
\label{ccos}
\ba
\widehat c_\ell(\hbar)&= -{2 \pi^2 \over 3}\widehat a_\ell (\hbar)+ 2\widehat a_\ell(\hbar) E_0^2  - \widehat b_\ell (\hbar) E_0 -\sum_{m \not= \ell} {\widehat a_m (2 \ell;\hbar) \over (m-\ell)^2}  + \sum_{m \not= \ell} {\widehat 
b_m (2 \ell;\hbar) \over 2(m-\ell)} \\
&+ 4 \CI_2(-2\ell) - \left({2 \pi^2 \over 3}-{\hbar^2 \over 48} \right) \CI_0(-2 \ell), \\
\widehat{d}_\ell(\hbar)&= -4\sum_{m\ge1 } {\widehat a_m (2\ell+1;\hbar) \over (2m-(2\ell+1))^2}  + \sum_{m \ge 1} {\widehat 
b_m (2\ell+1;\hbar) \over 2m-(2\ell+1)} + 4 \CI_2(-(2\ell+1)) \\
&- \left({2 \pi^2 \over 3}-{\hbar^2 \over 48} \right) \CI_0(-(2\ell+1)). 
\ea
\ee
Although these expressions look complicated, the derivatives of these quantities w.r.t. $E_0$ have a simple expression in terms 
of the perturbative quantum volume. This is because
\be
{\partial \over \partial E_0} \left(  {1\over 2 \pi \hbar}  \int_{E_0}^\infty {{\rm vol}_{\rm p} (E) \rd E \over \re^{E-\mu}+1} \right)=-{1\over 2 \pi \hbar}  {{\rm vol}_{\rm p} (E_0) \over \re^{E_0-\mu}+1}, 
\ee
and we deduce
\be
\label{eoders}
\ba
{\partial \widehat A(\hbar) \over \partial E_0}&= -{1\over 2 \pi \hbar} {\rm vol}_{\rm p}(E_0), \\
{\partial \widehat c_\ell (\hbar) \over\partial  E_0}&= -{\re^{2 \ell E_0} \over 2 \pi \hbar} {\rm vol}_{\rm p}(E_0), \\
{\partial \widehat d_\ell(\hbar) \over \partial E_0}&= -{\re^{(2 \ell +1) E_0}\over 2 \pi \hbar} {\rm vol}_{\rm p}(E_0),
\ea
\ee
where $E_0$ is regarded as an independent variable (in particular, independent of $\hbar$). We can use these expressions to obtain explicit formulae for the $\hbar$ expansion 
of $\widehat A (\hbar)$,  $\widehat c_\ell (\hbar) $, $\widehat d_\ell (\hbar)$ in terms of integrals of ${\rm vol}_n (E)$. This makes it possible to evaluate analytically 
the infinite sums appearing 
in (\ref{ak}) and (\ref{ccos}), order by order in $\hbar$.

Let us now consider the contribution to the grand potential coming from the non-perturbative part of the quantum volume. 
We expand the exponent of (\ref{sm-series}) and write
\be
{\rm vol}_{\rm np}(E)= \sum_{m=1}^\infty \sum_{\ell=0}^\infty s_{\ell, m} (k) (-1)^m  \re^{-\left({4 m \over k} + 2\ell\right) E},
\ee
where the coefficients $s_{\ell, m} (k)$ can be obtained from $s_{n}(k)$ and $\widehat a_{r}(\hbar)$, with $n\le m$, $r\le \ell$. 
By using now (\ref{rsigma}), with 
\be
\sigma= {4 m \over k}+ 2 \ell, \qquad m=1, 2, \cdots, \quad \ell =0, 1, \cdots, 
\ee
we find  
\be
\label{fjnp}
\ba
 {1\over 2 \pi \hbar}  \int_{E_0}^\infty {{\rm vol}_{\rm np} (E) \rd E \over \re^{E-\mu}+1}&=-{1\over 2 \hbar}   \sum_{m=1}^\infty \csc\left( {4 \pi m \over k} \right) 
 \sum_{\ell=0}^\infty s_{\ell, m} (k) (-1)^m   \re^{-\left({4 m \over k} + 2\ell\right) \mu} \\
&+ {1\over 4 \pi^2 } \sum_{n \ge 0} \left\{  \sum_{m=1}^\infty \sum_{\ell=0}^\infty { (-1)^m s_{\ell, m} (k)  \over 4m + k \left(2 \ell -n\right)}  \re^{-\left({4 m \over k} + 2\ell\right) E_0} \right\} 
(-1)^n \re^{-n \left(\mu-E_0\right)}. 
\ea
\ee
If our conjecture (\ref{sm}) is true, we can write the first line of (\ref{fjnp}) as
\be
\label{jws}
  \sum_{m=1}^\infty d_m(k) (-1)^m   \re^{-4 m \mu_{\rm eff}/k},
\ee
where $\mu_{\rm eff}$ is defined by the same equality as (\ref{Effi}), i.e. 
\be
\label{muffi}
\mu_{\rm eff}=   \mu +{\pi^2 k  \over 2} \sum_{\ell=1}^\infty a_\ell(k) \re^{-2 \ell \mu}.
\ee
In order to obtain the grand potential, we have to add to the integral (\ref{jp-mu}) the infinite series of corrections in (\ref{j-int}). Since 
\be
f'(n)= -{E'(n) \over \re^{E(n)- \mu}+1}, 
\ee
it is easy to see that, when expanded around $\mu \rightarrow \infty$, these corrections can only lead to a constant term, plus a series in $\re^{- \mu}$. In addition, if we evaluate these corrections perturbatively in $\hbar$, we find that $f^{(2r-1)}(0)$ is of order $\hbar^{2r-1}$.

We can now compare this calculation 
to the existing results on the perturbative grand potential of ABJM theory \cite{mp,hmo,hmo1,cm,hmo2,hmmo}
\be
\begin{split}
\label{jp-kn}
J(\mu,k) &= {2 \mu^3 \over 3 k \pi^2} + \left( {1\over 3k} + {k \over 24}\right) \mu+ A(k) + \sum_{\ell \ge 1} \left( a_\ell (k) \mu^2  + b_\ell (k) \mu  + c_\ell (k) 
 \right)\re^{-2 \ell  \mu} \\
 &+\sum_{m=1}^\infty d_m(k) (-1)^m   \re^{-4 m \mu_{\rm eff}/k}. 
 \end{split}
\ee
As we have seen, the correction terms in (\ref{j-int}) only affect the constant term and the series in $\re^{-\mu}$. In particular, the terms in $\mu^2 \re^{-2 \ell \mu}$ and $\mu \re^{-2 \ell \mu}$ in $J(\mu, k)$ are completely captured by the integral (\ref{jp-mu}). We conclude that:
\be
\label{abrels}
a_\ell (k)=-{ \widehat a_\ell (\hbar)\over \pi^2 k}~, \qquad b_\ell(k)= {\widehat b_\ell (\hbar)\over 2 \pi^2 k} . 
 \ee
 This is the main conjecture in \cite{hmmo}, which was tested there by direct computation. Here, we have derived this conjecture from the WKB solution of the spectral problem. Furthermore, (\ref{jws}) is exactly the non-perturbative contribution (in $k$) to the grand potential. It contains the contribution 
of worldsheet instantons, which is determined by the relation to standard topological string theory \cite{mpabjm,dmp,mp,hmo1}, as well as the contribution of 
bound states conjectured in \cite{hmo2}. In fact, 
our conjecture (\ref{sm}) was tailored to reproduce the worldsheet instanton contribution. Although we can not derive (\ref{jws}) from first principles 
(since we have not proved (\ref{sm})), the appearance of the ``effective" chemical potential $\mu_{\rm eff}$ 
is a simple consequence of the fact that instanton corrections in the WKB method 
involve the exponentiated quantum A-period (\ref{A-vm}), as it was already pointed out in \cite{mp}. 

The conclusion of this analysis is that the conjecture of \cite{hmmo} relating the $a$, $b$ coefficients of the grand potential to the refined topological string can be derived from 
the perturbative WKB analysis of the spectral problem, and the conjecture on the contribution of bound states in \cite{hmo2} can be also partially justified. Conversely, the known 
results about the worldsheet instanton contribution to the grand potential lead to the conjecture (\ref{np-vol}) for the non-perturbative correction to the WKB 
quantization condition. 

However, our analysis of (\ref{j-int}) leads to further results for the grand potential, since we have not yet considered the 
constant term and the series in $\re^{-\mu}$. Let us first look at the 
constant term. First of all, it is easy to see that the constant part (in $\mu$) of the series of corrections in (\ref{j-int}), as $\mu \rightarrow \infty$, is given by 
\be
\label{series-cons}
\sum_{r \ge 1} {B_{2r} \over (2r)!} E^{(2r-1)}(0), 
\ee
and involve the derivatives of the function $E(n)$ at $n=0$. We conclude that
\be
\label{aa}
A(k)=\widehat A(\hbar)+ {1\over 4 \pi^2 } \sum_{m=1}^\infty \sum_{\ell=0}^\infty { (-1)^m s_{\ell, m} (k)  \over 4m + 2 \ell k}  \re^{-\left({4 m \over k} + 2\ell\right) E_0} 
+ \sum_{r \ge 1} {B_{2r} \over (2r)!} E^{(2r-1)}(0). 
\ee
The second term in the r.h.s. comes from the contribution of $n=0$ appearing in (\ref{fjnp}), and it is non-perturbative in $\hbar$. We can still analyze this equality by looking 
at the perturbative expansion around $k=0$. The l.h.s. has a power series expansion of the form \cite{mp,hanada}
\be
\label{ak-ser}
A(k)= {2 \zeta(3) \over \pi^2 k}-{k \over 12} +\cdots
\ee
In order to test the equality of these two expressions, we can expand the first and third term in the r.h.s. of (\ref{aa}) in powers of $k$ and compare the resulting series order by order. Notice however that each order in (\ref{ak}) is given by an infinite sum. At leading order, the third term in (\ref{aa}) does not contribute, and in the expression (\ref{ak}) we can use the values of the coefficients (\ref{aobo}) as well as the leading order term for $n=0$ in (\ref{enseries}). If we plug in these values in (\ref{ak}), we reproduce (numerically) the first term in (\ref{ak-ser}), testing in this way the proposed equality at leading order. Alternatively, writing
\be
\widehat{A}(\hbar)=\sum_{n\geq 0}\widehat{A}_n k^{2n-1} 
\ee
we can use the first equation in (\ref{eoders}) and $\eqref{volexact}$ to obtain the expression
\be
\widehat{A}_0 =E_0 +\frac{2 \zeta(3)}{\pi^2}- \frac{\re^{E_0}}{8\pi^3}  G_{4,4}^{2,4}\left(\frac{\re^{2E_0}}{16}\left|
\begin{array}{c}
 \frac{1}{2},\frac{1}{2},\frac{1}{2},\frac{1}{2} \\
 0,0,-\frac{1}{2},-\frac{1}{2}
\end{array}
\right.\right)~,
\ee
where the constant of integration is fixed by matching the expansion $\eqref{ak}$. With this expression we can perform an analytic check\footnote{This analytic check has also been independently performed by Yasuyuki Hatsuda in an unpublished note.}. 

Similarly, we can now look at the coefficients in the expansion in powers of $\re^{-\mu}$ in both sides of (\ref{j-int}). The equality in (\ref{j-int}) 
implies that the coefficients $\widehat c_\ell (\hbar)$ are related to the coefficients $c_\ell(k)$ appearing in (\ref{jp-kn}) by the equation  
\be
\label{c-eq}
\ba
c_\ell (k)& ={\widehat c_\ell (\hbar)\over 2 \pi^2 k} +\frac{\re^{2 \ell E_0}}{4\pi^2} \sum_{m=1}^\infty \sum_{r=0}^\infty { (-1)^m s_{r, m} (k)  \over 4m + 2 k \left( r -\ell\right)}  \re^{-\left({4 m \over k} + 2r \right) E_0}
\\
&- \sum_{r \ge 1} {B_{2r} \over (2r)!}\left[ f^{(2r-1)}(0)\right]_{-2 \ell}, \qquad \ell \ge 1,
\ea
\ee
where the bracket $\left[ \cdot \right]_{-2 \ell}$ means the term in $\re^{-2\ell \mu}$ in the expansion of $ f^{(2r-1)}(0)$ at large $\mu$. 
Finally, we obtain the condition, 
\be
\label{d-eq}
\ba
{\widehat{d}_\ell(\hbar)\over 2\pi^2 k} & + \frac{\re^{(2 \ell +1)  E_0}}{4\pi^2} \sum_{m=1}^\infty \sum_{r=0}^\infty { (-1)^m s_{r, m} (k)  \over 4m +  k \left( 2r -2\ell-1\right)}  \re^{-\left({4 m \over k} + 2r \right) E_0} \\\ &+ \sum_{r \ge 1} {B_{2r} \over (2r)!} \left[ f^{(2r-1)}(0)\right]_{-2 \ell-1}=0, \qquad \ell \ge 0. 
\ea
\ee
As before, these equations can be tested in perturbation theory around $k=0$. In this case, the terms coming from the last sum in (\ref{fjnp}) do not contribute. In fact, the function $c_\ell (k)$ is 
known to be an analytic function of $k$ at the origin, therefore it should be given by the perturbative series around $k=0$ of the r.h.s., while the non-perturbative terms in $k$ in the r.h.s. should cancel. At leading order in $k$, the 
corrections involving $f^{(2r-1)}(0)$ do not contribute either, and the resulting equalities can be checked for the very first values of $\ell$, by performing numerically the sums or by using the 
equations (\ref{eoders}). 

It is instructive to perform a test of the various equalities we have written down, again in perturbation theory around $k=0$, but at next-to-leading order in $k$. This can be done in a single 
strike by looking at (\ref{j-int}) in perturbation theory. After taking into account the various expansions in $\hbar$, and after writing down\footnote{$J_{\rm p}(\mu,k)$ denotes the perturbative, in $k$, part of $J(\mu,k)$.}
\be
J_{\rm p}(\mu, k)= \sum_{n\ge 0} J_n (\mu) \hbar^{2n-1}, 
\ee
it is easy to see that a next-to-leading order test amounts to checking that 
\be
\label{ntl}
J_1(\mu)=\frac{1}{2\pi}\int_{E^{(0)}_0}^\infty \frac{ {\rm vol}_1(E)\rd E}{\re^{E-\mu}+1}-\frac{1}{96(1+ 4\, \re^{-\mu})}. 
\ee
The second term in the r.h.s. has two contributions: one coming from the expansion of $E_0^{\rm p}$ at order $\hbar^2$, 
where we have taken into account (\ref{ens}), and another from the 
term $r=1$ in the series of corrections in (\ref{j-int}). The integral appearing in the r.h.s. can be evaluated explicitly, as a power series in $\re^{\mu}$, by using the explicit 
result (\ref{vol01}). The result is, 
\be
\frac{1}{2\pi}\int_{E^{(0)}_0}^\infty \frac{ {\rm vol}_1(E)\rd E}{\re^{E-\mu}+1}=- \sum_{n\ge 1} \left( r_n  +s_n\right) (-1)^n \re^{n\mu}, 
\ee
where
\be
r_n=\frac{1}{3} 2^{-2 n-5}, \qquad s_n=-\frac{4^{-n-4} n \left(n^2-1\right) \Gamma \left(\frac{n}{2}\right)^2}{3\, \Gamma \left(\frac{n+3}{2}\right)^2}.
\ee
It is easy to see that
\be
-\sum_{n\ge 1} r_n (-1)^n \re^{n \mu}=\frac{1}{96(1+ 4\, \re^{-\mu})}, 
\ee
and also that 
\be
s_n= {Z^{(1)}_n\over n}, 
\ee
where 
\be
J_1(\mu)= -\sum_{n\ge 1} {Z^{(1)}_n \over n} \left( - \re^{\mu} \right)^n, 
\ee
and the explicit expression for $Z^{(1)}_n$ was derived in \cite{mp} from the Wigner--Kirkwood expansion and written down in eq. (5.20) of that paper. This proves (\ref{ntl}).

Although we have tested these equalities perturbatively, it would be also very interesting to test them non-perturbatively, for finite values of $k$, 
even numerically. In order to do this, one might need to perform a Borel resummation of the asymptotic series appearing in the Euler--Maclaurin expansion, following \cite{cg}. 

We conclude that our calculation of the quantum volume for the spectral problem (\ref{diff-eq}) proves, to a large extent, the conjectures made in \cite{hmmo} relating the 
membrane instanton part of the grand potential to the refined topological string in the NS limit. This is the content of our result (\ref{abrels}). 
The incorporation of ``bound states" by promoting $\mu$ to $\mu_{\rm eff}$, as proposed in \cite{hmo2}, follows also from our WKB analysis. 
For a complete derivation of the known results, one should also prove the equalities (\ref{aa}), (\ref{c-eq}) and (\ref{d-eq}). 
\sectiono{Conclusions and open problems}

In this paper we have analyzed in detail a spectral problem which appears in ABJM theory and in the theory of quantum spectral curves. 
This problem is a particular case of the quantization of the mirror curve of local $\IP^1 \times \IP^1$, 
for some particular values of the complex parameters $z_1, z_2$. This choice of parameters is very convenient since, 
after imposing appropriate analyticity conditions, 
it leads to a reformulation of the problem in terms of an integral equation (\ref{int-eq}) for which we 
have a lot of information --numerical and analytical-- thanks to its r\^ole in ABJM theory. 

Our main conclusion is that the WKB analysis of \cite{mm,acdkv} (which reproduces the known results for 
the refined topological string in the NS limit) is insufficient for actually calculating the spectrum: there are instanton effects 
which should be added, just as in other quantum-mechanical situations \cite{zj,zjj}. 
Surprisingly, the relevant instanton series is essentially the free energy of the {\it standard} topological string. 
This shows that the NS limit of the refined string (i.e. the choice $\epsilon_2=0$ in the Omega background) and the 
conventional topological string (i.e. the choice $\epsilon_1=-\epsilon_2$) are intimately related in a non-perturbative treatment. This was already shown in 
the ``dual" calculation of \cite{hmmo}, by using the non-perturbative definition provided by the lens space matrix model. 

Mathematically, the conjectural WKB quantization condition for the spectral problem (\ref{int-eq}) or (\ref{diff-eq}) is very surprising. In particular, it contains 
information about the Gopakumar--Vafa invariants of local $\IP^1 \times \IP^1$. It would be interesting to derive this quantization condition analytically, perhaps along 
the lines of \cite{flz}. The spectral problem for integral equations like (\ref{int-eq}) is also related to a TBA equation \cite{zamo,tw}, and it would be also 
interesting to understand the non-perturbative corrections from the point of view of the TBA. 

Our exact quantization condition is also very interesting from the point of view of the WKB method. Usually, the perturbative WKB quantization leads to an {\it asymptotic} series 
in the energy which has to be resummed with some appropriate prescription. Instanton corrections are typically needed when the series is not Borel-summable, and they 
cancel the non-perturbative ambiguities appearing in the resummation process, as in the double-well potential studied in \cite{zjj}. 
In the case studied here, the WKB quantization condition leads to an infinite series which seems to be {\it convergent} in the energy plane. However, its coefficients develop simple poles at values of $\hbar$ of the form $2 \pi n$, where $n$ is an integer. Instanton corrections are 
needed to cancel these non-physical poles, as in the closely related HMO mechanism of \cite{hmo2}. Therefore, the analytic mechanism for combining non-perturbative and perturbative corrections in our example is very different from 
standard quantum-mechanical situations: for generic, real values of $\hbar$, we have a WKB convergent series which requires however non-perturbative corrections. 

Of course, the main question for future research is the following: is this story an accident of this example, 
or is it a general feature of local CY manifolds? 
One possible strategy to answer this question
would be to look at the quantized curves of local CYs and take the spectral problem seriously. 
For example, the quantization of the mirror curve for local $\IP^2$ leads to the operator \cite{acdkv}:
\be
\label{localp2}
-1+ \re^{\hat u} + \re^{\hat v}+ z  \re^{-\hat u - \hat v}.
\ee
After a choice of polarization, we can require this operator to annihilate the wavefunction and obtain a difference equation. 
This equation can be analyzed as in \cite{acdkv}, where it 
was shown that the perturbative WKB periods give the standard 
refined topological string on local $\IP^2$, in the NS limit. However, this does not lead by 
itself to a quantized spectrum, since we need additional analyticity conditions on the wavefunction. One could impose a natural set of conditions similar to 
the ones we imposed on (\ref{diff-eq}) (see \cite{sergeev} for a related discussion). As we mentioned before, 
the quantum B-period is singular for infinitely many values of $\hbar$. 
If the difference equation defined by (\ref{localp2}) with the appropriate analyticity conditions leads to a well-defined spectral problem 
for these values of $\hbar$, then non-perturbative instanton corrections are needed. 
It might be the case that these corrections involve the standard topological string, as in the example studied in this paper. 

Our result suggests in fact a new point of view on the non-perturbative definition of topological string theory: start with a well-defined spectral problem arising 
in the quantization of the spectral curve, and find its exact WKB quantization condition. The perturbative part will be given by the refined topological string in the NS limit, 
and the non-perturbative corrections might lead to the Gopakumar--Vafa expansion of the usual topological string, just as in our example. 

This non-perturbative definition depends ultimately on an appropriate definition of the spectral problem. 
There is however a family of CY geometries in which this problem should be well-defined, thanks to the result of \cite{ns}: 
these are the $A_{N-1}$ local geometries, which generalize local $\IP^1 \times \IP^1$. In this case, the spectral problem is the Baxter equation obtained by quantizing the 
spectral curve of the relativistic Toda lattice. However, this equation is an auxiliary tool to solve the original quantum integrable system, and the analyticity properties of the 
solution to the Baxter equation should follow from the original problem of determining the spectrum of the conserved Hamiltonians for the quantum, relativistic Toda lattice 
(this is well understood in the non-relativistic limit, see for example \cite{pg,kl}). Interestingly, the $A_{N-1}$ geometries also have a dual large $N$ Chern--Simons 
description, as well as a matrix model description \cite{akmv}. It would be very interesting to study non-perturbatively the quantum spectral curves of these geometries 
and understand their relation to the matrix models. In the case studied in this paper, the matrix model partition function is the partition function of an ideal gas of $N$ fermions with the energy eigenvalues determined by the spectral problem.

Another possible extension of this paper is the analysis of the spectral problem appearing in the Fermi gas formulation of $\CN=3$ Chern--Simons--matter models. 
In \cite{mp}, generalized kernels were defined for many of these theories; it would be interesting to study their spectrum with the techniques developed here. 
This would lead to new results for non-perturbative effects in the type IIA/M-theory duals to these theories.  

We hope to report on these problems in the near future. 

\section*{Acknowledgements}
We would like to thank Sergei Lukyanov, Sanefumi Moriyama, Kazumi Okuyama and 
Samson Shatashvili for useful discussions and correspondence. We are particularly thankful to Yasuyuki Hatsuda, who 
took this paper seriously and found some errors in its first version. His comments prompted us to correct some of the formulae in section 4. 
J.K. would like to thank the Simons Center for Geometry 
and Physics and the 2013 Simons Summer Workshop where part of this work was carried out.
This work is supported in part by the Fonds National Suisse, subsidies 200020-137523 and 200020-141329. 

\appendix

\sectiono{Mellin transform}
We will use the Mellin transform as defined in (\ref{mt}). The inverse Mellin transform can be computed 
by inspection after using the following elementary integral:
\be
\label{inv-mt}
\int_0^1 u^{n-s-1} \left( \log u \right)^m \rd u = -{\Gamma(m+1) \over (s-n)^{m+1}}.
\ee
Regular terms in $s$ do not contribute to $\mu$ expansion of the inverse Mellin transform. Therefore, we first compute the Mellin transform, consider the polar part 
of the Laurent series, and perform the inverse Mellin transform by using (\ref{inv-mt}). 

We will do a Mellin transform of quantities in the grand canonical ensemble with respect to the variable
\be
u= \re^{-\mu}. 
\ee
The Mellin transform of the Fermi occupation number is 
\be
\label{01int}
\int_0^1 {u^{-s-1} \over \re^E u +1} \rd u = \re^{sE} I(s) + \sum_{k \ge 1} {(-1)^k \re^{-k E} \over s+k}, 
\ee
where 
\be
\label{is}
I(s)= \int_0^\infty {x^s \over 1+x} \rd x = - \pi \csc (\pi s). 
\ee
Notice that the function in the r.h.s. of (\ref{01int}) does not have poles at negative integer values of $s$. 
The function $I(s)$ has simple poles at the positive integers. The Laurent series around even non-negative integers is, 
\be
I(s)=-{1\over s -2\ell}-{\pi^2 \over 6} (s-2 \ell )+ \cdots
\ee
while for odd positive integers we find, 
\be
 I(s)={1\over s -(2\ell+1)}+{\pi^2 \over 6} \left(s-(2 \ell+1) \right )+ \cdots~.
\ee
Using these results, it is easy to compute the Mellin transform of $J_{\rm p}(\mu, k)$:
\be
\label{jp-mt}
\ba
\widehat J_{\rm p} (s, k)= {1 \over 2\pi^2 k}& \biggl\{ 4 \CI_2(-s)-\left( {2 \pi^2 \over 3} - {\hbar^2 \over 48} \right) \CI_0(-s) 
\\ &  \qquad - 4 \sum_{\ell \ge 1} \widehat a_\ell(\hbar) \CI_1(2\ell-s) +  \sum_{\ell \ge 1} \widehat b_\ell(\hbar) \CI_0 (2\ell-s)  \biggr\} I(s)+\cdots, 
\ea
\ee
where $\CI_j(n)$ are defined in (\ref{cijn}), 
and the dots in (\ref{jp-mt}) denote terms which are regular in $s$ or come from the second term in (\ref{01int}). 
To implement the inverse Mellin transform, we notice that this function has a pole of order fourth at $s=0$, with Laurent expansion
\be
{4 \over \pi^2 k}{1\over  s^4} +\left( {1\over 3 k  }  + {k \over 24} \right) {1\over s^2}- {\widehat A(\hbar) \over s}+\cdots
\ee
where $\widehat A(\hbar)$ is given in (\ref{ak}). At even positive integers, we have triple poles with the Laurent expansion
\be
{2\over \pi^2 k } {\widehat a_\ell (\hbar)  \over (s- 2 \ell)^3} + {1\over 2 \pi^2 k } {\widehat b_\ell (\hbar)  \over (s- 2 \ell)^2} - {1\over 2 \pi^2 k } {\widehat c_\ell (\hbar) \over s- 2 \ell}+\cdots
\ee
where $\widehat c_\ell$ is given in the first line of (\ref{ccos}). At odd positive integers, we have simple poles with the structure
 \be
{1\over 2 \pi^2 k } {\widehat{d}_\ell(\hbar) \over s-(2\ell+1)}~,
\ee
where $\widehat{d}_\ell$ is given in the second line of (\ref{ccos}). Putting all the results together, we derive the result in (\ref{jp-mt}).

Equivalently, one can use the Mellin transform to calculate the integrals (\ref{rjm}). Their values are easily found to be, 
\be
\ba
R^{(0)}_\ell (E_0, \mu)&=\sum_{ n=1, \atop n\neq 2\ell}{ \frac{(-1)^n}{2\ell-n}\re^{-(2\ell-n) E_0}\re^{-n\mu}}+\left(\mu-E_0\right)\re^{-2\ell \mu}+\frac{\re^{-2\ell E_0}}{2\ell},\\
 R^{(1)}_\ell(E_0, \mu)&=\sum_{n=1, \atop  n\neq 2\ell}{(-1)^n \left(\frac{1}{(2\ell-n)^2}+\frac{E_0}{2\ell-n}\right)\re^{-(2\ell-n) E_0}\re^{-n\mu}}\\
& +\left(\frac{\pi^2}{6}+\frac{\mu^2-E_0^2}{2}\right)\re^{-2\ell\mu} +\frac{\left(1+2\ell E_0\right)}{(2\ell)^2}\re^{-2\ell E_0}.
\ea
\ee
These expressions are valid for real $\mu> E_0$. Finally, we compute the integral (\ref{r-sig}). The Mellin transform is
\be
\widehat R_{\sigma}(E_0, s)= \CI_0( \sigma-s) I(s) + \cdots. 
\ee
If $\sigma$ is not an integer, we have two types of simple poles in this expression: poles at $s=\sigma$ coming from $\CI_0(\sigma-s)$, and poles coming from $I(s)$ at integer 
values $s=n$ (like before, there are no poles at negative integers). A simple calculation shows that the inverse Mellin transform is given by
\be
\label{rsigma}
R_{\sigma}(E_0, \mu)= -\pi \csc ( \pi \sigma)\re^{-\sigma \mu} + \sum_{n \ge 0} {(-1)^n \over \sigma-n} \re^{-(\sigma-n) E_0} \re^{-n \mu},
\ee
which again is valid for real $\mu> E_0$. 
Notice that the first term is what we would obtain with the Sommerfeld expansion of 
the Fermi factor used in for example \cite{wilson}, and the second term is a non-perturbative correction. 
This result can be also written down in terms of the Hurwitz Lerch transcendent $\Phi(a,b,c)$, 
\be
R_{\sigma}(E_0, \mu)= -\pi \csc( \pi \sigma)\re^{-\sigma \mu} + {1\over \sigma} \re^{-\sigma E_0} + \re^{E_0(1-\sigma)-\mu} \Phi\left( \re^{E_0-\mu}, 1, 1-\sigma\right). 
\ee
\sectiono{Higher order quantum volume} \label{qvolder}
In this appendix we will sketch how to derive the result $\eqref{vol01}$. It is obtained by solving equation $\eqref{diff-eq}$ with the ansatz $\eqref{WKBpsi}$ and then performing the period integral $\eqref{p-vol}$. When solving $\eqref{diff-eq}$ we will follow the approach given in \cite{dingle}.

Let us write the difference equation $\eqref{diff-eq}$ as 
\be
\left[\re^{\frac{\ri \hbar}{2}\partial_x}+\re^{-\frac{\ri \hbar}{2}\partial_x}\right]\psi(x)=2 r(x)\psi(x)
\ee
where
\be
r(x)=\frac{\re^{E}}{4\cosh{\left(\frac{x}{2}\right)}}~.
\ee
Plugging in the ansatz $\eqref{WKBpsi}$ in the above equation and expanding around $\hbar=0$ keeping $S(x,\hbar)$ fixed we get the following equation:
\be
\cosh{\left(\frac{1}{2}S^{(1)}(x,\hbar)-\frac{\hbar^2}{2^3}\frac{S^{(3)}(x,\hbar)}{3!}+\ldots\right)}\exp{\left(-\frac{\ri \hbar}{2^2}\frac{S^{(2)}(x,\hbar)}{2!}+\ldots\right)}=r(x)~,
\ee
where
\be
S^{(m)}(x,\hbar)=\partial_{x}^m S(x,\hbar)~.
\ee
We notice that a derivative of $S(x,\hbar)$ of order $m$ always comes with a factor $\hbar^{m-1}$, $m>0$. If we plug in the expansion $\eqref{WKBS}$ we can therefore first solve for $\partial_x S_0$ and then find $\partial_x S_n(x)$, $n>0$, in terms of derivatives of $\partial_x S_m(x)$ with $m<n$. For the first three orders we find
\be
\begin{split} \label{Sn}
\partial_x S_0(x)&=2\cosh^{-1}\left[r(x)\right]=p(x)~, \\
\partial_x S_1(x)&=\frac{\ri r(x)  S_0^{(2)}(x)}{4\sqrt{r(x)^2-1}}~, \\
\partial_x S_2(x)&=\frac{r(x)}{64 \left(r(x)^2-1\right)^{3/2}}\left[\left( S_0^{(2)}(x)\right)^2+16\left(r(x)^2-1\right) S_1^{(2)}(x)\right]+\frac{ S_0^{(3)}(x)}{48}~.
\end{split}
\ee
The perturbative part of the quantum volume is given by $\eqref{p-vol}$, and it has an expansion for small $\hbar$ given in $\eqref{vol-p-exp}$. In order to find ${\rm vol}_n(E)$ we use $\eqref{Sn}$ and calculate the integral. To lowest order we find the result $\eqref{volexact}$. The order $\hbar$ contribution vanishes as it should and at order $\hbar^2$ we obtain the result $\eqref{vol01}$.

\end{document}